\documentclass[a4paper,10pt]{article}    
\pdfoutput=1
\usepackage[OT1]{fontenc} %Computer Modern
\usepackage{multicol}
\usepackage{graphicx}  
\usepackage{bbold} 
\usepackage{mathtools}
\usepackage{amsmath} 
\numberwithin{equation}{section} %equation numbering 
\usepackage{amsfonts} 
\usepackage{amssymb}
\usepackage{amsbsy}
\usepackage{amsfonts}
\usepackage{physics}
\usepackage{bbold} 
\usepackage{slashed}   
\usepackage{color}
\usepackage[table]{xcolor} 
\usepackage{tablestyles}
\usepackage{tabularx}
\usepackage{hyperref}
\usepackage{orcidlink}
\usepackage{bbding} 
\usepackage{tcolorbox}
\usepackage{ulem}

\definecolor{oucrimsonred}{rgb}{0.6, 0.0, 0.0}
\definecolor{DarkGray}{gray}{0.4}
\definecolor{forestgreen}{rgb}{0.13,0.35,0.13}
\definecolor{ocre}{HTML}{F16723}

\usepackage[page,toc,titletoc,title]{appendix}
 
\numberwithin{equation}{section}
\numberwithin{table}{section}
\numberwithin{figure}{section}
\usepackage{bm}% bold math
\usepackage{cite}  %  to make successive citations hyphened
\usepackage{mathrsfs}
\usepackage{framed}
\usepackage[font=footnotesize,labelfont=bf]{caption}
\def\eq#1{{Eq.~(\ref{#1})}}

\def\Tr{\mbox{Tr}\,}

\newcommand{\di}{\mbox{d}}

\def\di{\mbox{d}}

\colorlet{grayline}{gray!70}
\definecolor{blueline}{rgb}{0,0.27,0.55}
\definecolor{DarkGray}{gray}{0.4}
\definecolor{Gray}{gray}{0.6}
\definecolor{oucrimsonred}{rgb}{0.6, 0.0, 0.0}
\definecolor{persianblue}{rgb}{0.11, 0.22, 0.73}
\definecolor{forestgreen}{rgb}{0.13,0.35,0.13}
 \hypersetup{colorlinks, citecolor=forestgreen, linkcolor=forestgreen, urlcolor=forestgreen}
\newcommand{\be}{\begin{equation}}
\newcommand{\ee}{\end{equation}}
\newcommand{\bea}{\begin{eqnarray}}
\newcommand{\eea}{\end{eqnarray}}
\newcommand{\nn}{\nonumber}

\newcommand{\CC}{\operatorname{C}}
\newcommand{\BB}{\operatorname{B}}
\newcommand{\hk}{{\bf k}}
\newcommand{\hr}{{\bf r}}
\newcommand{\hn}{{\bf n}}
\newcommand{\hp}{{\bf p}}

\newcommand{\lambdap}{\lambda^{\prime}}

\newcommand*\xbar[1]{%
  \hbox{\;%
    \vbox{%
      \hrule height 0.5pt % The actual bar
      \kern0.5ex%         % Distance between bar and symbol
      \hbox{%
        \kern-0.25em%      % Shortening on the left side
        \ensuremath{#1}%
        \kern-0.07em%      % Shortening on the right side
      }%
    }%
  }%
} 
\newcommand{\com}[1]{}
\newcommand{\gsim}{\lower.7ex\hbox{$\;\stackrel{\textstyle>}{\sim}\;$}}
\newcommand{\lsim}{\lower.7ex\hbox{$\;\stackrel{\textstyle<}{\sim}\;$}} 

\newcommand{\bc}{\begin{center}}
\newcommand{\ec}{\end{center}}
\newcommand{\K}{K^{*}(892)^0}

\font\beeg=cmr17 scaled 1800
\newbox\ibox
\def\versal#1{\setbox\ibox=\hbox{{\beeg #1}~}%
	    \noindent\global\hangindent=\wd\ibox\global\hangafter-2%
	    \sc\smash{\llap {\lower 14pt \box\ibox}}}
\begin{document}

\hypersetup{citecolor = forestgreen,
linktoc = section, 
linkcolor = forestgreen, 
urlcolor = forestgreen
}

%\onecolumn
\thispagestyle{empty}

\begin{tcolorbox}[colframe=gray!7,colback=gray!7]
\bc
{ \Large \color{oucrimsonred} \textbf{ 
 Tests of quantum contextuality in particle physics}}
 \ec
\end{tcolorbox}

\bc
\vspace*{1.5cm}
{\color{DarkGray}
  {\bf M. Fabbrichesi$^{a\, \orcidlink{0000-0003-1937-3854}}$,}
{\bf   R. Floreanini$^{a\, \orcidlink{0000-0002-0424-2707}}$,}
{\bf E. Gabrielli$^{b,a,c\, \orcidlink{0000-0002-0637-5124}}$ and} 
 {\bf L. Marzola$^{c,d\, \orcidlink{0000-0003-2045-1100}}$}
}\\

\vspace{0.5cm}
{\small 
{\it  \color{DarkGray} (a)
INFN, Sezione di Trieste, Via Valerio 2, I-34127 Trieste, Italy}
\\[1mm]
  {\it \color{DarkGray}
    (b) Physics Department, University of Trieste, Strada Costiera 11,  I-34151 Trieste, Italy}
  \\[1mm]  
  {\it \color{DarkGray}
(c) Laboratory of High-Energy and Computational Physics, NICPB, R\"avala 10, 10143 Tallinn, Estonia}
 \\[1mm]
  {\it \color{DarkGray}
(d) Institute of Computer Science, University of Tartu, Narva mnt 18,  51009 Tartu, Estonia.}
}
\ec

 \vskip0.5cm
\bc
{\color{DarkGray}
\rule{0.7\textwidth}{0.5pt}}
\ec
\vskip2cm
  \begin{tcolorbox}[colframe=gray!7,colback=gray!7]
\bc
{\bf ABSTRACT} 
\ec
\noindent  Quantum contextuality  refers to the impossibility of assigning  a predefined, intrinsic value to a physical property of a system independently of  the context in which the property is measured. It is, perhaps, the most fundamental feature of quantum mechanics. The many states with different spin that particle physics  provides are the ideal setting for  testing contextuality. We verify that the polarization states of single spin-1 massive particles produced at colliders are contextual. 
We test $W^{+}$ gauge bosons produced in top-quark decays, $J/\psi$ and $\K$ mesons in $B$-meson decays and $\phi$ mesons in  $\chi^0_c$ and $\chi^1_c$ charmonium decays by reinterpreting  the data and the analyses of the ATLAS, LHCb, Belle II and BESIII experimental collaborations, respectively. The polarization states of these four particles show contextuality with a significance larger than $5\sigma$. We also discuss the presence of quantum contextuality in spin states of bipartite systems formed by spin-1/2 particles. We test $\Lambda$ and $\Sigma$ baryons reinterpreting two BESIII data analyses, and pairs of top quarks utilizing a recent analysis of the CMS collaboration.  Quantum contextuality is present with a significance exceeding $5\sigma$  also in these cases. In addition, we study the feasibility of testing quantum contextuality by means of $Z$ boson production in association with the Higgs boson, $Z$ and $W$ bosons pairs created in Higgs boson decays and with pairs of $\tau$ leptons. For the latter, we use Monte Carlo simulations that mimic the settings of SuperKEKB and of future lepton colliders. Experiments at high energies, though not designed for the purpose, perform surprisingly well in testing for quantum contextuality.
\end{tcolorbox}
\vspace*{1cm}

  \vskip 3cm
\bc 
{\color{DarkGray} %\vbox{$\bowtie$} 
%\SixFlowerPetalDotted 
\SquareShadowBottomRight
}
\ec

%%%%%%%%%%%%%%%%%%%%%%%%%%%%%%%%%%%%%%%%%%%%%%%%%%%%%%%
	\newpage

	\tableofcontents
	
	%to begin the line numbers: 
	%\linenumbers
	
	%%%%%%%%%%%%%%%%%%%%%%%%%%%%%%%%%%%%%%%%%%%%%%%%%%%%%%%%%%%%%%%%%%%%%%%%%%%%%%%%%%%%%%%%%%%%%%%%%%%%%%%%%%%%%%%%%%%%%%%%%%%%%%%

\newpage
\section{Introduction\label{sec:intro}} 

{\versal  Even though particle physics} is based on quantum field theory, and therefore incorporates quantum mechanics, it has been at least partially shielded from the quirkiest aspects of the theory by the $S$-matrix formalism, within which most experimental results are typically discussed. This is no longer true as an increasing number of works  (see~\cite{Barr:2024djo} for a review article) have come to address directly several features of quantum mechanics observable in collider experiments---among them, \textbf{entanglement}~\cite{Horodecki:2009zz,Benatti:2010,Nielsen:2012yss,bruss2019quantum} 
and \textbf{Bell nonlocality}~\cite{Bell:1964,scarani2019bell} and, now, contextuality.

Take $\cal A$, $\cal B$ and $\cal C$ to be three observables, and assume that $\cal A$ commutes with $\cal B$ and with $\cal C$, but $\cal B$ does not commute with $\cal C$. Hence, according to the principles of quantum mechanics, it is possible to measure $\cal A$ along with $\cal B$, or $\cal A$ along with $\cal C$, but one cannot measure simultaneously $\cal A$, $\cal B$ and $\cal C$---that is, in a single experimental run.  Would the value obtained for $\cal A$ be the same when measured with $\cal B$ as when measured with $\cal C$? If the answer is `Yes'  then quantum mechanics (or whatever theory one is using to analyze this situation) is \textbf{non-contextual}. Differently, if the answer is `No', or at least `No' in certain cases, quantum mechanics (or the theory in question) is \textbf{contextual}. The measurement outcome for $\cal A$ then depends upon whether it is measured along with $\cal B$, 
in the $({\cal A},{\cal B})$ context, or together with $\cal C$, in the $({\cal A},{\cal C})$ context.

\textbf{Quantum contextuality} can be perhaps considered the most characteristic property of quantum mechanics. It combines different distinctive manifestations of quantum theory in a single
framework: from the impossibility of simultaneously measuring arbitrary
observables, to Bell nonlocality and entanglement. The Bell-Kochen-Specker theorem~\cite{Bell:1964fg,Kochen:1968zz} shows that a context-independent classical descriptions of the predictions
of quantum theory is indeed impossible---at least for Hilbert spaces of dimension larger than two. 
On general grounds, since any measurement implies an interaction between the system being measured and the measuring apparatus,
it should not be surprising that the result of the observation could depend on the state of the system as well as on that of the measuring device~\cite{Bell:1964fg}. Quantum contextuality has been experimentally tested at low energies  by means of photons~\cite{PhysRevLett.84.5457,PhysRevLett.90.250401,PhysRevLett.103.160405,Lapkiewicz:2011rpt,PhysRevLett.109.150401,PhysRevX.3.011012,Ahrens2013}, neutrons~\cite{nature425,PhysRevLett.103.040403}, ions~\cite{Kirchmair2009,PhysRevLett.110.070401,PhysRevLett.120.180401,doi:10.1126/sciadv.abk1660}, molecular nuclear spins~\cite{PhysRevLett.104.160501}, superconducting systems~\cite{Jerger2016} and nuclear spins~\cite{PhysRevLett.123.050401}. 
A more in-depth introduction to the topic, as well as a comprehensive list of references, can be found in two recent reviews~\cite{thompson2013,Budroni:2021rmt}.
 
\textbf{Hidden-variable models} (see, for instance, the review article~\cite{Genovese:2005nw}) must necessarily entail the same degree of contextuality as quantum mechanics, as well as of nonlocality---as implied by the experimental results on the violation of the Bell inequality. When combined, these results make any such model unpalatable and, necessarily, rather far from a realistic description of physical objects---which was their original motivation.

The simplest systems bound to show a degree of contextuality are those described by a Hilbert space of dimension three. An explicit example is provided by the spin state of \textbf{massive spin-1 particles}. The category includes elementary particles like the massive gauge bosons mediating the weak interaction, as well as composite states such as the spin-1 mesons produced in the decays of $B$-meson and charmonium states. All these particles are
copiously produced at collider experiments, their polarization studied in detail, and the related data analyses are available in the literature. We then reinterpret the data analyses of the ATLAS, LHCb, Belle II and BESIII experimental collaborations to, respectively, test quantum contextuality with $W^{+}$ gauge bosons produced in top-quark decays,  $J/\psi$ and $\K$ mesons resulting from  $B$-meson decays, as well as with $\phi$ mesons generated in $\chi^0_c$ and $\chi^1_c$ charmonium decays. All these four cases show contextuality with a significance of more than $5\sigma$.

\textbf{Bipartite systems of spin-1/2 particles} are also routinely produced at colliders. We show the presence of quantum contextuality in a Hilbert space of dimension four by reinterpreting the BESIII data analysis of the spin state of $\Lambda$ and $\Sigma$ baryon pairs. We also investigate contextuality with the top and anti-top quark pairs produced at the LHC.

Alongside these reinterpretations of experimental analyses, we also present analytical results for the $W$ and $Z$ gauge bosons, top quark and  $\tau$-lepton pairs---relying also on a Monte Carlo simulation for the latter---which probe the significance achievable in dedicated experimental tests of contextuality.

To facilitate the reading, all non-contextuality results based on a reinterpretation of experimental data are highlighted with boxes, to distinguish them from the estimates resulting from analytic or Monte Carlo computations that aim to show the feasibility of the measurement and its possible significance.

Throughout the work we utilize \textbf{quantum state tomography} to access the polarization matrices of the systems under consideration and study entanglement and contextuality. A discussion of the advantages and of the limitations of this technique can be found in~\cite{Fabbrichesi:2025aqp}. 

The results presented in the paper demonstrate the study and establish the detection of contextuality at high energies and in the presence of strong and electroweak interactions. Particle physics provides the ideal setting for these studies and the ultimate testing ground for the predictions of quantum mechanics.

%%%%%%%%%%%%%%%%%%%%%%%%%%%%%%%%%%%%%%%%%%%%%%%%%%%
\section{Non-contextuality inequalities}

{\versal In physics, contextuality } describes how, or whether, the details of an observation affect what is observed. The result of a measurement could depend on how the measurement was made, as well as on the specific combination of observables that we chose to measure simultaneously. Based on these ideas, contextuality implies the impossibility of assigning intrinsic values to observables that transcend the specifics of the experimental protocol. In other words, contextuality states the impossibility of characterizing a physical property independently of what other properties---the context---are simultaneously measured with it. 

Non-contextual hidden variable models, instead, attribute to physical observables set values which exist independently of the context provided by the simultaneous measurement of additional observables. In general, this requirement does not preserve the mutual algebraic relations among observables predicted by quantum mechanics, thereby allowing for measurable experimental consequences. Contextuality is expected to be present in all quantum systems described by Hilbert spaces of dimension three~\cite{Bell:1964fg,Kochen:1968zz} or larger. Accordingly, we can investigate quantum contextuality  in particle physics by using single spin-1 massive particles, which spin state (a qutrit) belongs to an Hilbert space of dimension three. Next by complexity, we can use composite systems formed by two spin-1/2 particles, a bipartite qubit system described by the tensor product of two dimension-two Hilbert spaces.

\subsection{Three-level  systems}

\begin{figure}[ht!]
\begin{center}
\includegraphics[width=3in]{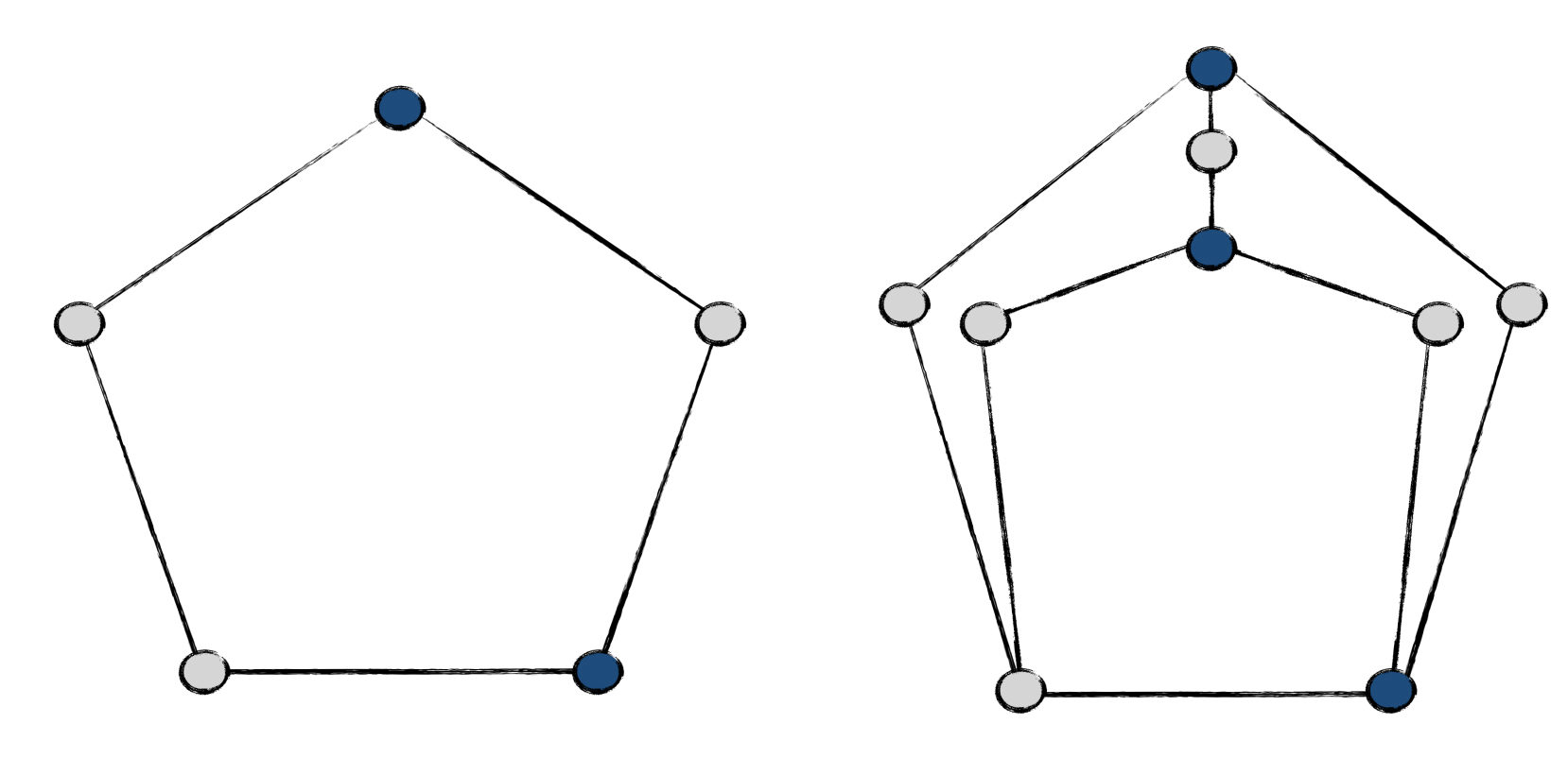}
 \caption{\footnotesize Graphs of five and nine projector operators showing the classical exclusivity conditions. The dark vertices correspond to the possible values of 1 assigned in a context independent manner.  A gray vertex represents the  value 0.
\label{fig:chains} 
}
\end{center}
\end{figure}

Testing quantum contextuality with qutrits involves five dichotomic observables ${\cal O}_i$, $i=1,2,3,4,5$, each with output
$o_i$ taking the values $\pm 1$, chosen so that the operator ${\cal O}_i$ commutes with ${\cal O}_{i+1}$, but in general not with the remaining ones. These observables are to be measured in pairs, as specified by the grouping
$\{ (1,2),\ (2,3),\ (3,4),\ (4,5),\ (5,1) \}$ and, since in a non-contextual
theory the outcomes $o_i$ have well-defined values, the following inequality holds:
\begin{equation}
o_1\,o_2\, + \, o_2\,o_3\, + \, o_3\,o_4\, + \, o_4\,o_5\, + \, o_5\,o_1\, \geq -3\ .
\label{inequality}
\end{equation}
Equivalently, the corresponding inequality for the expectation values of the observables is given by:
\begin{equation}
\langle{\cal O}_1\,{\cal O}_2\rangle\, + \, \langle{\cal O}_2\,{\cal O}_3\rangle\, 
+ \, \langle{\cal O}_3\,{\cal O}_4\rangle\, + \, \langle{\cal O}_4\,{\cal O}_5\rangle\, 
+ \, \langle{\cal O}_5\,{\cal O}_1\rangle\, \geq -3\ .
\label{quantum_inequality}
\end{equation}
A graphical representation of this relation can be obtained by introducing a set of three-dimensional projectors
\be
\Pi_i \equiv | v_i \rangle \langle v_i |, 
\label{eq:projectors}
\ee
specified here by the vectors $|v_i\rangle$, $ i=1,\dots, 5\ ,$ with $\langle v_i |v_{i+1}\rangle=\,0$ so that the observables ${\cal O}_i = 1-2\,\Pi_i$ respect the above assignments. Each vector $|v_i\rangle$ can then be associated to the vertex of a pentagon, as in Fig.(\ref{fig:chains}), and the inequality (\ref{quantum_inequality}) reduces to
\begin{equation}
\sum_{i=1}^5 \langle \Pi_i \rangle \leq 2\ .
\label{inequality_5}
\end{equation}
As we shall explicitly see, it is possible to choose five binary and pairwise-commuting observables or, equivalently, five pairwise-orthogonal vectors such that the inequality (\ref{inequality_5}) is violated when tested on a qutrit state implemented with a massive spin-1 particle.

This construction can be generalized by using more observables or, equivalently, more vectors organized as vertices of graphs~\cite{PhysRevLett.112.040401} more elaborated than a simple pentagon. In general, the larger the number of operators involved in the non-contextuality inequality, the larger is the amount of states that do violate it. It has been shown that the minimum number of projector operators needed for constructing a non-contextuality 
test violated by all spin-1 quantum states is 13~\cite{Cabello_2016}. 
Qutrit states produced at colliders can be tested to determine the minimum number of operators necessary to make contextuality apparent. While a pure state should violate a non-contextual inequality associated with a graph possessing just 5 vertices~\cite{Klyachko:2008zz}, all the states beside fully incoherent mixtures will exhibit contextuality when a test based on a 9 vertices graph is utilized~\cite{PhysRevA.86.042125}. A qutrit state given by a fully incoherent mixture only shows contextuality if a graph containing 13 vertices is employed~\cite{PhysRevLett.108.030402}.

The projection operators used in \eq{eq:projectors}, as well as those used in tests with 9 and 13 operators, can be given explicitly by specifying sets of unit vectors that, pairwise only, satisfy the usual orthonormality conditions.

The set of 5 unit vectors in \eq{eq:projectors} can be taken to be~\cite{ahrens2013fundamentalexperimentaltestsnonclassicality}
\be
|v_j\rangle  = \Big( \cos \phi, \sin \phi \, \cos \frac{4 \pi j}{5}, \sin \phi \, \sin \frac{4 \pi j}{5} \Big)^T \quad \text{with} \quad \cos^2 \phi = \frac{\cos \pi/5}{1 + \cos \pi/5}\, ,
\ee
with $j=1,2,3,4$ and 5.
The set of 9 unit vectors, instead, is given by~\cite{PhysRevA.86.042125}
\begin{align}
|v_1 \rangle  & = (1,0,0)^T,\quad  &|v_2 \rangle  & = (0,1,0)^T, \quad & |v_3 \rangle  & = (0,0,1)^T,  \nn \\
|v_4 \rangle  & =1/\sqrt{2} (0,1,-1)^T,\quad  &|v_5 \rangle  & = 1/\sqrt{3} (1,0,-\sqrt{2})^T, \quad & |v_6 \rangle  & = 1/\sqrt{3} (1,\sqrt{2},0)^T, \nn  \\
|v_7 \rangle  & = 1/2(\sqrt{2},1,1)^T,\quad  &|v_8 \rangle  & = 1/2 (\sqrt{2},-1,-1)^T, \quad & |v_9 \rangle  & = 1/2 (\sqrt{2},-1,1)^T,
\end{align}
while the 13 unit vectors are finally given as~\cite{PhysRevLett.108.030402}
\begin{align}
|v_1 \rangle  & = (1,0,0)^T,\quad  &|v_2 \rangle  & =1/\sqrt{2} (0,1,1)^T, \quad & |v_3 \rangle  & = 1/\sqrt{3}(1,-1,1)^T,  \nn \\
|v_4 \rangle  & =1/\sqrt{2} (1,1,0)^T,\quad  &|v_5 \rangle  & = (0,0,1)^T, \quad & |v_6 \rangle  & = 1/\sqrt{2} (0,1,-1)^T, \nn  \\
|v_7 \rangle  & = 1/\sqrt{3} (1,1,1)^T,\quad  &|v_8 \rangle  & = 1\sqrt{2}(1,-1,0)^T, \quad & |v_9 \rangle  & = 1/\sqrt{2} (1,0,-1)^T,  \nn\\
|v_{10} \rangle  & = 1/\sqrt{2} (1,0,1)^T,\quad  &|v_{11} \rangle  & = (0,1,0)^T, \quad & |v_{12} \rangle  & = 1/\sqrt{3} (-1,1,1)^T,  \nn\\
 & & |v_{13} \rangle  & = 1/\sqrt{3} (1,1,-1)^T\, . & &
\end{align} 
Alternative choices are discussed in the review articles~\cite{thompson2013,Budroni:2021rmt}.
%%%%%%%%%%%%%
%%%%%%%%%%%% 

Non-contextuality is verified if the sum of the expectation values of the $N$ projection operators satisfies the following inequality:
\be
\boxed{\mathbb{CNTXT}_N  \equiv  \sum_{i=1}^N \langle  \Pi_i  \rangle= \sum_{i=1}^N \Tr (\rho\,  \Pi_i) \leq c_N\,,} \label{eq:def}
\ee
in which $\rho$ is the qutrit density matrix, $c_5 = 2$, $c_9=3$ and $c_{13} = 4$. These values are found by considering the largest number of possible insertions of projectors with the same outcome~1 in the corresponding graph, without violating the condition that two consecutive vertices 
cannot have assigned the value 1~\cite{PhysRevLett.112.040401}, as shown, for the case of 5 and 9 operators, in Fig.~\ref{fig:chains}. The simplest case with a 5 operator chain is equivalent to the Klyachko-Can-Binicioglu-Shumovsky  (KCBS) inequality~\cite{Klyachko:2008zz}. The test using a chain of 9 projectors was introduced in Ref.~\cite{PhysRevA.86.042125}, while the case with 13 operators was used by S.\ Yu and C.H.\ Oh in their proof of the Bell-Kochen-Specker theorem, which was originally based on 117 unit vectors~\cite{Kochen:1968zz}. In quantum mechanics, the inequalities in~\eq{eq:def} have an upper limit as well: it is $\sqrt{5}$ for the KCBS case, 10/3 for 9 operator tests and 13/3 for the 13 operator chain.

The non-contextuality inequality in \eq{eq:def} is a `Yes' or `No' condition and, as such, does not lend itself to a quantitative estimate. Nevertheless, the significance of the answer---the confidence in, say, the `No' answer---does depend on the numerical value of $\mathbb{CNTXT}_N$.  The result of the contextuality test can be optimized through a unitary matrix $V$ implicitly defined by the condition $[\Pi', \rho]=0$, with $\Pi' = V\Pi V^\dagger$ and $\Pi=\sum_i \Pi_i$~\cite{PhysRevA.86.042125}. The trace operation is then to be taken after the eigenvalues of the two matrices are ordered by size and matched by rearranging the eigenvectors in $V$, yielding
\be
\overline{\mathbb{CNTXT}}_N = \sum_{k=1}^{4} \lambda^\rho_k \, \lambda^\Pi_k\,  \leq c_N\,, \label{eq:max}
\ee
in which $\lambda_i^\rho$ are the eigenvalues of the matrix $\rho$  and $\lambda_i^\Pi$ those of the matrix $\Pi$.

The test using $\mathbb{CNTXT}_{13}$ is special since the sum of the projection operators is proportional to the identity matrix. Consequently, the test becomes independent of the state used in the computation of the expectation values. However, in order to actually test the inequality $\mathbb{CNTXT}_{13} <4$, it is necessary to experimentally probe directly the operator algebra---which is impossible at current collider detectors. Since in quantum state tomography experimental data are used to reconstruct the polarization matrices of the considered processes, only contextuality inequalities that are state dependent can be used.

In the applications we discuss in Sec.~\ref{sec:test-spin-1}, the violation of the non-contextuality inequality will come to depend on one or more of the kinematic parameters that regulate the production process of the quantum system of interest. The choice pertaining to which values of these parameters are to be considered in the inequality is then relevant when assessing the significance of the violation---which we do by taking benchmark values.

\subsection{Four-level systems}
 %%%%%%%%%%%%%%%%%%%%%%%%%%%%%%%
 
The next case we analyze is that of a system described by an Hilbert space of dimension four. Within particle physics, this is best exemplified by a bipartite system formed by two spin-1/2 particles.

Let us consider nine dichotomic observables, organized into a $3\times 3$ matrix ${\cal O}_{ij}$, $i,j=1,2,3$, with corresponding outputs $o_{ij}$ that take the two values $\pm 1$. We then consider the following combination of outputs: 
\be
o_{11}\, o_{12}\, o_{13} + o_{21}\, o_{22}\, o_{23} +o_{31}\, o_{32}\, o_{33} + o_{11}\, o_{21}\, o_{31} 
+ o_{12}\, o_{22}\, o_{32} - o_{13}\, o_{23}\, o_{33}\ ,
\label{PM-combination}
\ee
involving products of the rows and of the columns of the matrix $o_{ij}$. In a non-contextual theory, the nine outcomes $o_{ij}$ have predefined assignments; when all of them are 1, the combination (\ref{PM-combination}) reaches its maximum value of 4: changing the assignment of just
one output $o_{ij}$ to $-1$ changes the signs of two of the addenda in the combination (\ref{PM-combination}),
thereby reducing the total result to less than 4.

Assuming that the observables ${\cal O}_{ij}$ belonging to the same row or  column are simultaneously measurable, from the previous considerations one obtains the following inequality for the expectation values\cite{Cabello2008,Budroni:2021rmt}:
\be
\langle {\cal O}_{11}\, {\cal O}_{12}\, {\cal O}_{13}\rangle + \langle{\cal O}_{21}\, {\cal O}_{22}\, {\cal O}_{23}\rangle 
+ \langle{\cal O}_{31}\, {\cal O}_{32}\, {\cal O}_{33}\rangle + \langle{\cal O}_{11}\, {\cal O}_{21}\, {\cal O}_{31}\rangle 
+ \langle{\cal O}_{12}\, {\cal O}_{22}\, {\cal O}_{32}\rangle - \langle{\cal O}_{13}\, {\cal O}_{23}\, {\cal O}_{33}\rangle
\leq 4\ .
\label{PM-inequality}
\ee
For bipartite qubit system, a convenient choice of observables ${\cal O}_{ij}$ involves the Pauli matrices $\sigma_i$, $i=1,2,3$: 
\begin{align}
{\cal O}_{11}&=\sigma_3\otimes \mathbb{1} &  {\cal O}_{12}&=\mathbb{1}\otimes \sigma_3
& {\cal O}_{13}&=\sigma_3\otimes\sigma_3\nn \\
{\cal O}_{21}&=\mathbb{1}\otimes \sigma_1 &  {\cal O}_{22}&=\sigma_1\otimes \mathbb{1} & {\cal O}_{23}&=\sigma_1\otimes\sigma_1 \nn \\
{\cal O}_{31}& =\sigma_3\otimes\sigma_1  & {\cal O}_{32}&=\sigma_1\otimes\sigma_3  &  {\cal O}_{33}        &=\sigma_{2}\otimes\sigma_{2} \,.\label{assignments}
\end{align}
Inserting these assignments in the expectation values of \eq{PM-inequality}, the inequality is violated independently of the state on which is to be evaluated. It is known as the Peres-Mermin square~\cite{Mermin:1990dqo,Peres:1990bjx,Peres:1991fff}. 

As in the spin-1 case, we are more interested in contextuality inequalities that are state dependent. These are obtained in the following manner.
The non-contextuality condition  in \eq{PM-inequality} can be reduced to a simpler, state dependent~\cite{Cabello2008,Nambu}, one by choosing $ {\cal O}_{33} =\mathbb{1}\otimes\mathbb{1}$ instead. With this choice, \eq{PM-inequality} becomes
\be
\boxed{\mathbb{CNTXT'} \equiv \big\langle(\sigma_3 \otimes \sigma_1)(\sigma_1 \otimes \sigma_3)\big\rangle -
\big\langle(\sigma_3 \otimes \sigma_3)(\sigma_1 \otimes \sigma_1)\big\rangle \leq 0}\label{nambu}
\ee
corresponding to the expectation value of the matrix
\be
\begin{pmatrix}
0 & 0 &0 &-2\\
0&0&2&0\\
0&2&0&0\\
-2&0&0&0\\
\end{pmatrix}\ .
\label{CNTXT-matrix}
\ee
The two averages in \eq{nambu} can take the values $\pm1$ when evaluated on the 
eigenvectors of (\ref{CNTXT-matrix}). The contextuality condition arises from the necessity of having the expectation values of the operator $(\sigma_3 \otimes \sigma_1)(\sigma_1 \otimes \sigma_3)$ positive while simultaneously that of the operator  $(\sigma_3 \otimes \sigma_3)(\sigma_1 \otimes \sigma_1)$ negative---which is impossible without context dependence. In quantum mechanics, the largest value possible for $\mathbb{CNTXT'} $ is 2.

Interestingly, if one instead chooses for the observables ${\cal O}_{12}$, ${\cal O}_{22}$, ${\cal O}_{31}$,
${\cal O}_{32}$ and ${\cal O}_{33}$ the same assignment $\mathbb{1}\otimes\mathbb{1}$ in place of those in (\ref{assignments}),
the non-contextual inequality (\ref{PM-inequality}) reduces to
\be
\langle {\cal O}_{11}\, {\cal O}_{13}\rangle + \langle{\cal O}_{21}\, {\cal O}_{23}\rangle 
+ \langle{\cal O}_{11}\, {\cal O}_{21}\rangle 
- \langle{\cal O}_{13}\, {\cal O}_{23}\rangle
\leq 2\ ,
\label{CHSH-inequality}
\ee
which is equivalent to the Clauser-Horn-Shimony-Holt Bell inequality~\cite{CHSH_1969,scarani2019bell}. This  shows a strict connection between contextuality and Bell non-locality supported by quantum theory for this class of systems~\cite{Cabello2021}.%
\footnote{A comprehensive way of treating together contextuality and non-locality is through the mathematics of sheaf theory~\cite{Abramsky_2011}.}

The inequality \eq{nambu}, being state-dependent, is violated only by a certain class of two-qubit states. Other complementary non-contextuality tests can be obtained by changing the reference Cartesian axes.
For instance, the two simplest rotations corresponding to the cyclic permutations of the Pauli matrices give
rise to the following inequalities:
\be
 \mathbb{CNTXT''} \equiv \big\langle(\sigma_1 \otimes \sigma_2)(\sigma_2 \otimes \sigma_1)\big\rangle -
\big\langle(\sigma_1 \otimes \sigma_1)(\sigma_2 \otimes \sigma_2)\big\rangle \leq 0\label{sigmas2}
\ee
and
\be
\mathbb{CNTXT'''} \equiv  \big\langle(\sigma_2 \otimes \sigma_3)(\sigma_3 \otimes \sigma_2)\big\rangle -
\big\langle(\sigma_2 \otimes \sigma_2)(\sigma_3 \otimes \sigma_3)\big\rangle \leq0 \, . \label{sigmas3}
\ee
The above inequalities can be used with the original one in~\eq{nambu} in testing quantum contextuality. More generally, we can maximize the non-contextuality violation by rotating the Pauli matrices in~\eq{nambu} at each point in the kinematic space by means of two unitary matrices, $U$ and $V$:
\be
(U \otimes V)^{\dag}\cdot \Big[(\sigma_3 \otimes \sigma_1)(\sigma_1 \otimes \sigma_3)-
(\sigma_3 \otimes \sigma_3)(\sigma_1 \otimes \sigma_1)\Big] \cdot (U \otimes V) \, . \label{eq:UV}
\ee
We will use this optimization when we come to discuss the non-contextuality inequalities for top-quark pairs.

As for the spin-1 case, the confidence in the answer of the proposed contextuality test depends on the numerical value  obtained. In Section 4 we study benchmark values  meant to maximize the significance of the violation when the non-contextuality inequalities depend on one, or more, kinematic parameters.

%%%%%%%%%%%%%%%%%%%%%%%%%%%%%%%%%%%%%%%%%%%%%%%%%%%
\section{Testing spin-1 massive particles} 
\label{sec:test-spin-1}

{\versal The simplest physical systems} in which contextuality can be tested need to be at least three-dimensional,
as the polarization space of massive spin-1 particles is.
Within the Standard Model (SM), the vector gauge bosons are the only massive spin-1 elementary particles, while additional 
massive states with spin-1 like the mesons are composite. Experimental data are available on $W$ gauge bosons and several spin-1 mesons---among them we shall consider the $J/\psi$, $\K$ and $\phi$. 
Reinterpreting the data provided by the experiments, the polarization or the helicity density matrices of such spin-1 states can be completely reconstructed: they can be used to estimate the expectation value of the projection operators in \eq{eq:projectors}. These matrices are either directly those describing the spin-1 particle or those for their production in association with some other particle. In the latter case, a partial trace over the other particle gives the desired density matrix.

The polarization of massive spin-1 particles can be conveniently described by $3\times 3$ polarization matrices that
can be expanded as
\com{%
\be
\rho_{_{1}} = \dfrac{1}{3}\, \mathbb{1}_{3\times 3} +\sum_{1}^{8} f_a T^a \ ,
\ee
in terms of the Gell-Mann matrices $T^a$ and the identity,
\be
\rho_{_{1}} =  \begin{pmatrix} \frac{1}{3} +f_{3} +\frac{f_{8}}{\sqrt{3}}& f_{1}-i f_{2} & f_{4}- i f_{5} \\
 f_{1}+i f_{2} &\frac{1}{3} -f_{3} +\frac{f_{8}}{\sqrt{3}} & f_{6}-i f_{7} \\
 f_{4}+i f_{5}& f_{6}+i f_{7}&\frac{1}{3} - 2 \frac{f_{8}}{\sqrt{3}}
\end{pmatrix}\ . \label{eq:rhoT}
\ee
Equivalently, one can also write:}
\be
\rho_{_{1}} = \dfrac{1}{3} \,\mathbb{1}_{3\times 3} + \dfrac{1}{2} \sum_{M=-1}^{+1} s_M S_M + \sum_{M=-2}^{+2} t_M T_M\ ,
\ee
in which $S_{\pm1} = \mp (S_x\pm i S_y)/\sqrt{2}$ and $S_0=S_z$ are the spin operators in the spherical basis and $T_M$ five rank 2 tensors with components
\be
T_{\pm 2} = S_{\pm1}^2\,, \quad T_{\pm 1} = \frac{1}{\sqrt{2}} ( S_{\pm 1} S_0 + S_0 S_{\pm 1} )\, ,\quad \text{and} \quad T_0 = \frac{1}{\sqrt{6}} ( S_{+1} S_{-1} + S_{-1} S_{+1} + 2 S_0^2 )\,.
\ee
The polarization of a massive spin-1 particle is therefore given by eight parameters: 3 linear and 5 tensor polarizations.
The polarization density matrix can then be explicitly expressed as
\be
\rho_{_{1}} =  \begin{pmatrix} \dfrac{1}{3} +\dfrac{s_{3}}{2}  +\dfrac{t_{0}}{\sqrt{6}} & -\dfrac{ s_{1}}{2}-\dfrac{t_{1}}{\sqrt{2}} & t_{2} \\
-\dfrac{ s_{-1}}{2}-\dfrac{t_{-1}}{\sqrt{2}} &\dfrac{1}{3} -\dfrac{2\, t_{0}}{\sqrt{6}} & -\dfrac{ s_{1}}{2}+\dfrac{t_{1}}{\sqrt{2}}  \\
t_{-2}& -\dfrac{ s_{-1}}{2}+\dfrac{t_{-1}}{\sqrt{2}} &  \dfrac{1}{3} -\dfrac{s_{3}}{2}  +\dfrac{t_{0}}{\sqrt{6}}
\end{pmatrix}. \label{eq:rhoTensor}
\ee

Since we are interested in assigning a significance to the violation of the non-contextuality inequalities in \eq{eq:def}, we need to propagate  the uncertainty in the polarization or helicity density matrices coming from the experimental analyses. We propagate the errors in these experimental inputs by constructing an ensemble of spin-1 states---described by  properly normalized density matrices---varying their matrix elements within the given uncertainties. 

An abridged version of this Section, containing a selection  of the results presented below, appeared in~\cite{Fabbrichesi:2025ifv}.

%%%%%%%%%%%%%%%%%%%%%%%%%%%%%%%%%%%%%%%%%%%
\subsection{$W$ gauge bosons in  $t\to W^+b$  decays}
%%%%%%%%%%%%%%%%%%%%%%%%%%%%%%%%%%%%%%%%%%%%%%%%%%%

The $3\times 3$ polarization density matrix  of the $W^+$ in the $t\to W^+ b$ decay, in the standard
helicity
basis $\{+1,0,-1\}$, is given in the Standard Model (SM) by~\cite{Aguilar-Saavedra:2010ljg}
\be
\left\{ 
\begin{array}{ccc}
\rho_{_{00}} &\propto& m_{_{00}}\, (1+ \cos \theta )\\
\rho_{_{++}}&=& 0\\
\rho_{_{--}} &\propto& m_{_{--}}\, (1- \cos \theta )\\
\rho_{_{+-}}=\rho_{_{-+}} & =& 0\\
\rho_{_{0+}} = \rho_{_{+0}} &=&0\\
\rho_{_{0-}} = \rho_{_{-0}} &\propto& m_{_{0-}} \, \sin \theta\, e^{i \varphi}, ,
\end{array}
\right. \label{eq:Wpol}
\ee
in which the angles $\theta$ and $\varphi$ define the direction of the spin of the top quark with respect of the direction of its momentum. The same is expected for $W^{-}$ in the decay $\bar t \to  W^- \bar b$. According to the SM, we have that
\bea
m_{_{00}} &=& \frac{m_t^2-m_W^2}{m_W^2} %= 3.611\pm 0.016
\, ,\nn\\
 m_{_{--}}&= &2\, \frac{m_t^2-m_W^2}{m_t^2} %=1.566\pm 0.002
 \, , \label{eq:msm}\\
 m_{_{0-}}&=& \sqrt{2} \,  \frac{m_t^2-m_W^2}{m_W \,m_t}%= 2.378 \pm 0.006
 \, ,\nn
\eea
in which $m_t$ and $m_W$ are the top quark and $W$ boson mass, respectively. The polarization matrix $\rho_W$ turns out to be a projector---that is $(\rho_W)^2=\rho_W$---showing explicitly 
that the $W$ gauge boson is in a pure state of polarization. The diagonalization of this polarization matrix gives, accordingly, a  matrix with 1 in one entry, zeros in the other two and no dependence on the input parameters in \eq{eq:msm}. The reason behind the purity of the $W^+$ state is that in the SM the coupling of the top quark to the charged gauge vector is chiral.

The pure state of the polarization of the $W^+$  used in the non-contextuality inequalities in \eq{eq:max} is factorized out and  the expectation values in the inequalities  give back the largest  possible values allowed by quantum mechanics, namely
\be
\overline{\mathbb{CNTXT}}_5=2.236 \quad \text{and} \quad \overline{\mathbb{CNTXT}}_9 = 3.333 \, .
\label{central}
\ee
 
\subsubsection{Using the polarization fractions}

We  use  the SM result to guide us in using the partial results of the experimental analysis. The polarization fractions of the $W$ gauge boson are defined as
\be
F_{i} \equiv \frac{\Gamma_{i} (t\to Wb)}{\Gamma_T(t\to Wb)}\, ,\quad \text{with}\quad  i=0,+,-
\ee
in which $\Gamma_{i}$ and $\Gamma_T$ are the corresponding polarized and total widths. They can be measured from  the decays of pairs of top quarks produced at the LHC in proton-proton collisions at a center-of-mass energy of $\sqrt{13}$ TeV. The measurement is performed selecting $t\bar t$ events decaying into final states with
two charged leptons (electrons or muons) and at least two $b$-tagged jet. The ATLAS Collaboration finds~\cite{ATLAS:2022rms}
$F_0 = 0.684 \pm 0.005|_{\text{stat}} \pm 0.014|_{\text{syst}}$, $F_- = 0.318 \pm 0.003|_{\text{stat}} \pm 0.008|_{\text{syst}}$ and  $F_+= -0.002\pm 0.002|_{\text{stat}}\pm 0.014|_{\text{syst}}$.

%%%%%%%%%%%%%
\begin{figure}[h!]
\begin{center}
\includegraphics[width=2.8in]{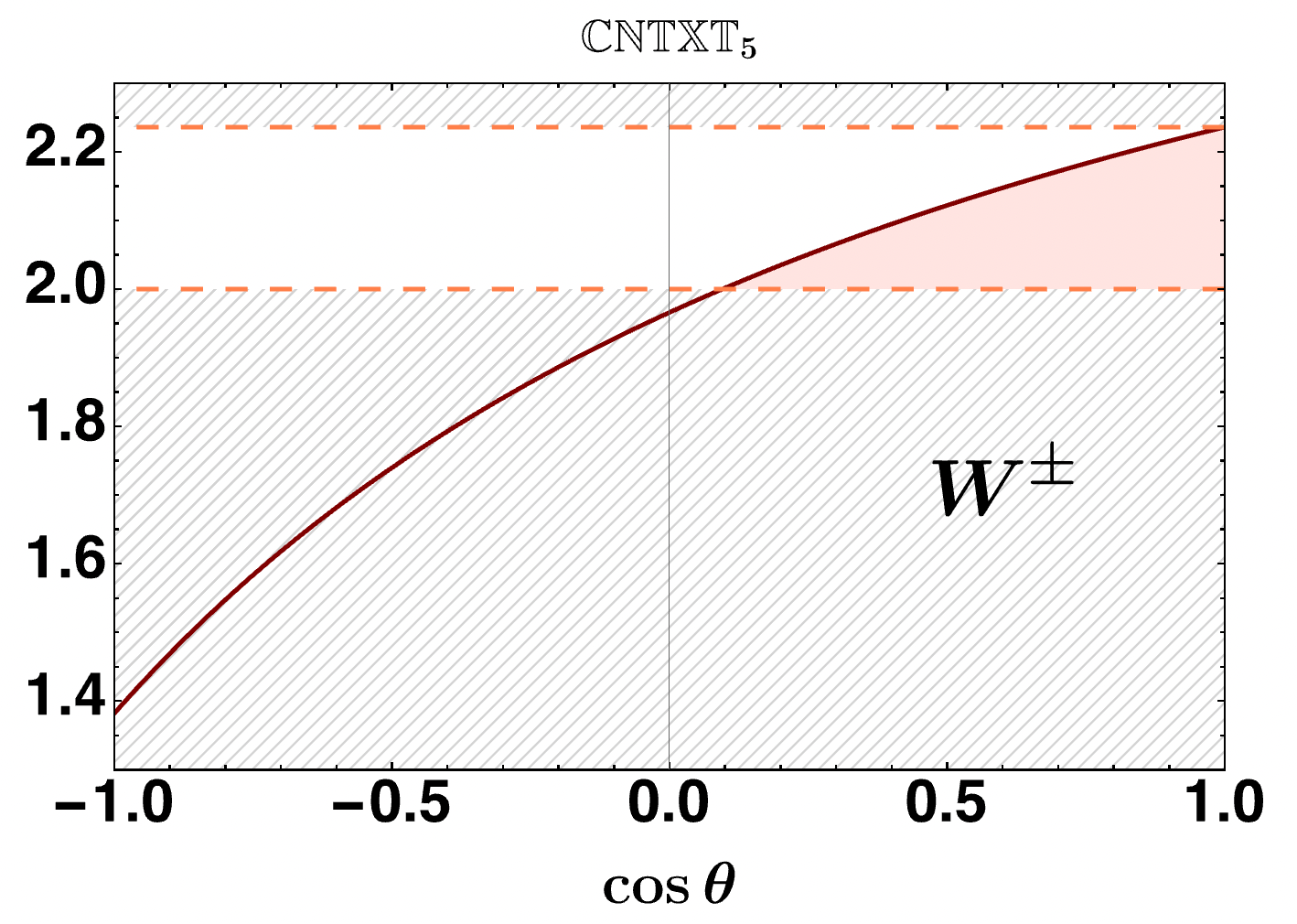}
\includegraphics[width=2.8in]{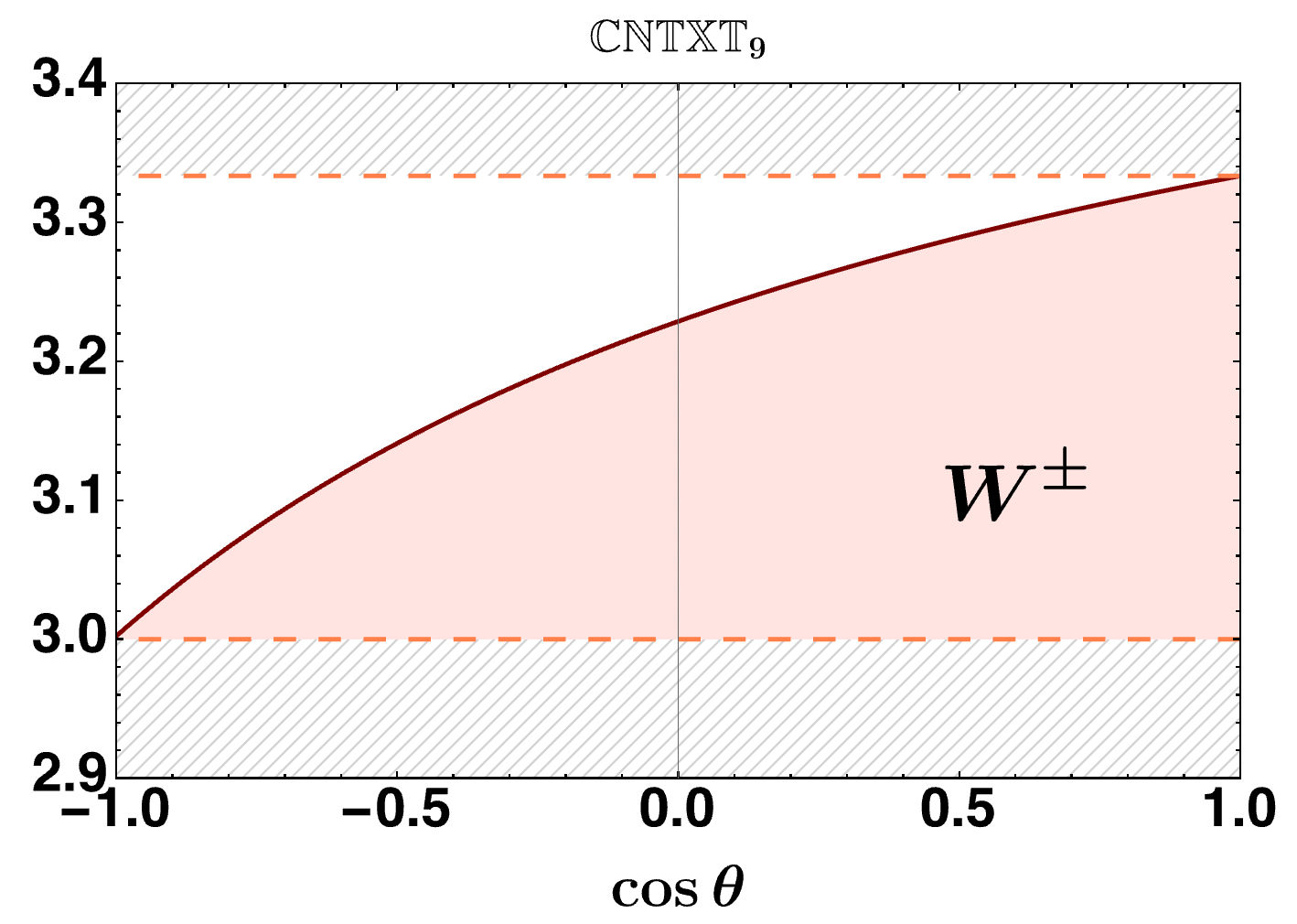}
 \caption{\footnotesize The curves plot $\mathbb{CNTXT}_5$ and  $\mathbb{CNTXT}_9$ for the $W$ gauge boson as a function of the cosine of the angle $\theta$ defined by the direction of the top-quark momentum and that of its spin. The horizontal, dashed lines mark, respectively,  the maximal value achievable assuming non-contextuality (lower line)  and the upper limit for contextuality within quantum mechanics (upper line).
\label{fig:context_W} 
}
\end{center}
\end{figure}
%%%%%%%%%%%% 

The coefficients $m_{_{00}}$, $m_{_{--}}$ and  $m_{_{++}}$ (though the latter  vanishes in the SM, we set it  to its experimental value)  are given in terms of the helicity fractions as
\bea
\frac{m_{_{00}}}{m_{_{00}}+m_{_{--}}+m_{_{++}}} &=& F_{0} =  0.684 \pm 0.015\, ,\nn \\ 
 \frac{m_{_{--}} }{m_{_{00}}+m_{_{--}}+m_{_{++}}}&=& F_-=0.318 \pm 0.009\, ,\nn\\
 \frac{ m_{_{++}}}{m_{_{00}}+m_{_{--}}+m_{_{++}}}&=&F_+= -0.002\pm 0.014\label{rho_exp}\, ,
\eea
with $F_0+F_-+F_+=1$; we have combined statistical and systematic uncertainties in quadrature. The values given by the helicity fractions are in agreement within $2\sigma$ with those from the SM, which are
\bea
\frac{m_{_{00}}}{m_{_{00}}+m_{_{--}}+m_{_{++}}} &=& \frac{m_{t}^{2}}{m_{t}^2+2 m_{W}^{2}}=0.698\pm0.005\, ,\nn \\ 
 \frac{m_{_{--}} }{m_{_{00}}+m_{_{--}}+m_{_{++}}}&=&  \frac{2 m_{W}^{2}}{m_{t}^2+2 m_{W}^{2}}=0.302\pm0.005\, ,\nn\\
 \frac{ m_{_{++}}}{m_{_{00}}+m_{_{--}}+m_{_{++}}}&=&0\, .
\eea

The lack of experimental data on the off-diagonal terms in \eq{rho_exp} makes it impossible to diagonalize the polarization matrix and maximize the violation of the non-contextuality bound. We use instead the polarization matrix as it is, diagonalizing only the projector sum so that the unknown off-diagonal terms do not contribute to the result of \eq{eq:def}. The values we obtain are then lower bounds for the non-contextuality of the involved particles.

The polarization density matrix for the $W^-$ is given by
\be
\rho_W \propto 
\begin{pmatrix} 
  m_{_{++}} (1+\cos\theta)& 0 &  0 \\
  0& m_{_{00}}  (1+\cos\theta) & m_{_{0-}}  \,\sin\theta \,e^{-i \varphi} \\
  0&m_{_{0-}} \,\sin\theta\, e^{i \varphi}& m_{_{--}} (1-\cos\theta)
  \end{pmatrix}
\label{eq:rhoW}
\ee
barring a normalization factor and with the values of the involved coefficients obtained from~\eq{rho_exp}.

Figure~\ref{fig:context_W} shows the values of $\mathbb{CNTXT}_5$ and  $\mathbb{CNTXT}_9$ for a sample of $W$ produced in top quark decays, computed with the central values in \eq{rho_exp}, as function of the cosine of the angle $\theta$.   Whereas the non-contextuality bound is exceeded only in the forward direction for $\mathbb{CNTXT}_5$, it is for all angles in the case of  the less restrictive $\mathbb{CNTXT}_9$.

Taking the angle $\theta=\pi/4$ as a benchmark value, we find 
\be
\boxed{\mathbb{CNTXT}_5=2.175\pm 0.015 \quad \text{and} \quad
 \mathbb{CNTXT}_9 = 3.311 \pm  0.012}\label{CNTXT_W}
 \ee
and a violation of the contextuality inequality with a significance well above $5\sigma$ for both the operators.  

The contextuality test above has been performed by partially using the SM  result about the coupling between the top and the $W$ gauge boson. A test completely independent of the SM will be possible as soon as the experimental collaborations provide data to reconstruct the complete polarization matrix.

%%%%%%%%%%%%%%%%%%%%%%%%%%%%%%%%%%%%%%%%%%%
\subsection{$Z$ gauge bosons in  the process $e^+e^-\to Z H$}
%%%%%%%%%%%%%%%%%%%%%%%%%%%%%%%%%%%%%%%%%%%%%%%%%%%

%%%%%%%%%%%%%
\begin{figure}[ht!]
\begin{center}
\includegraphics[width=2.5in]{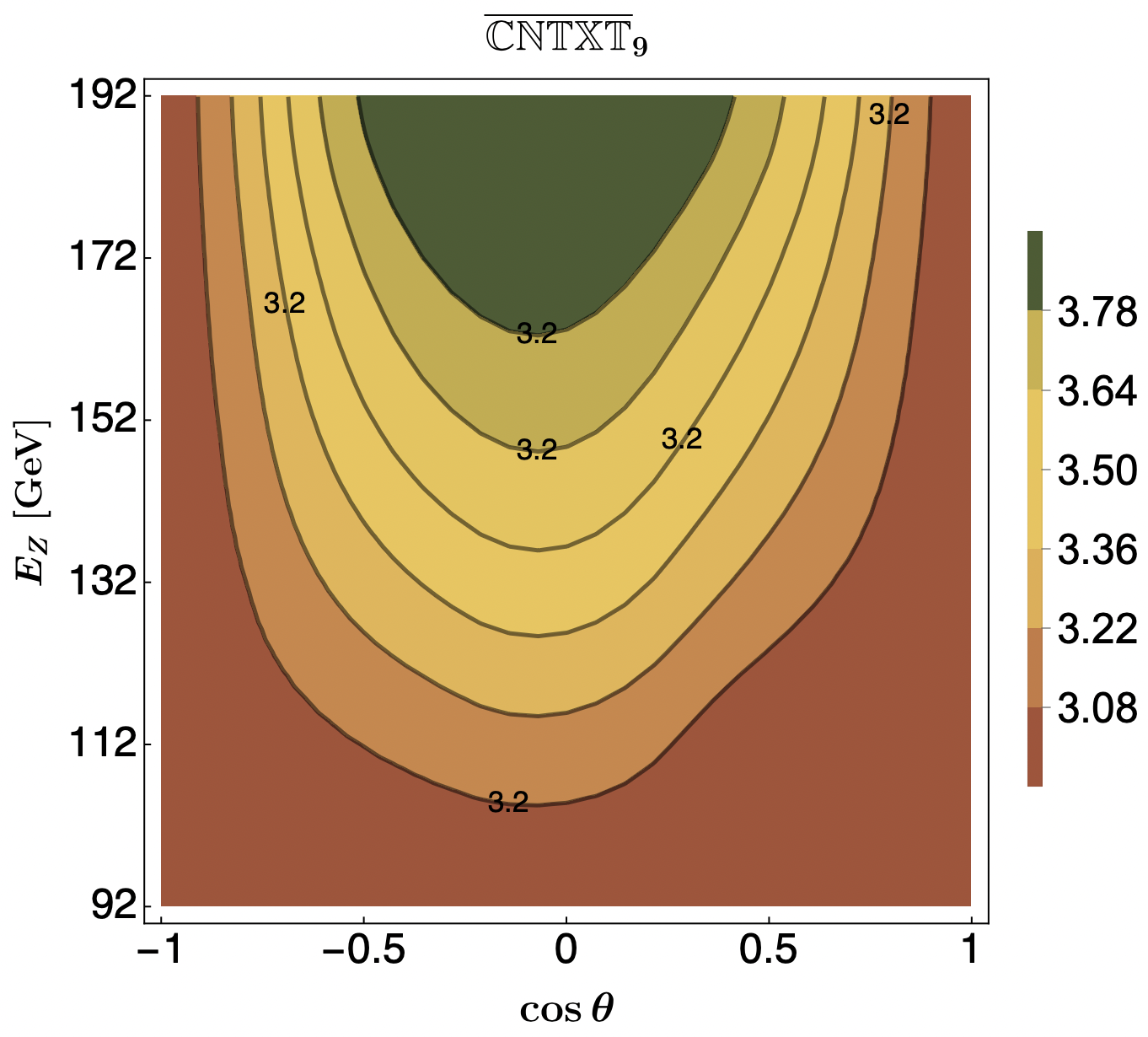}
\includegraphics[width=2.5in]{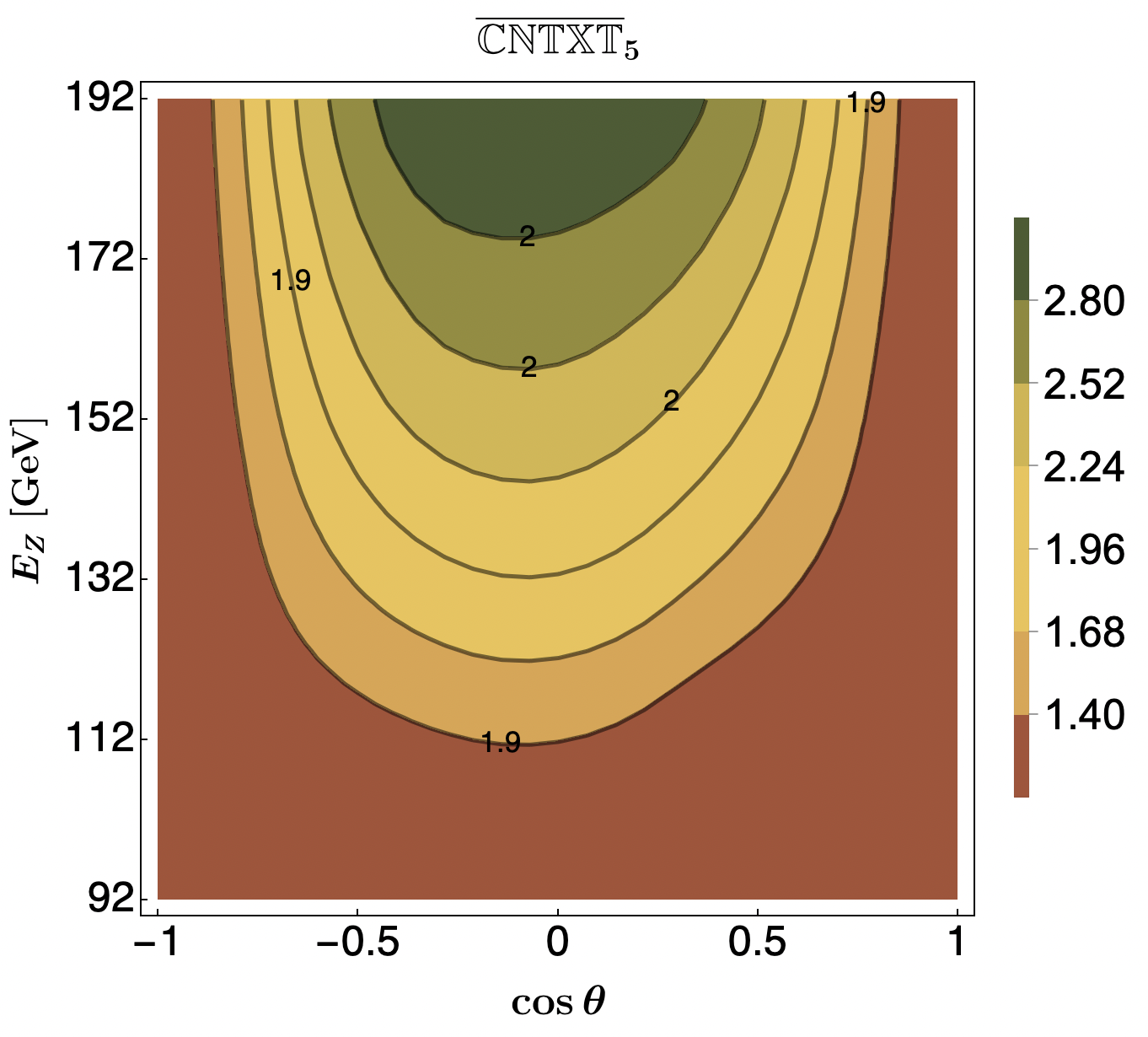}
 \caption{\footnotesize  $\overline{\mathbb{CNTXT}}_5$ and $\overline{\mathbb{CNTXT}}_9$ for the $Z$ gauge bosons produced in $e^+e^- \to HZ$ as a function of the cosine of the angle $\theta$  and its energy $E_{Z}$.
\label{fig:context_Z} 
}
\end{center}
\end{figure}
%%%%%%%%%%%% 

The other fundamental spin-1 particle in the SM is the neutral gauge boson. Unfortunately there are yet no data on the polarization of the $Z$ boson. Nevertheless, we can  study the analytic values of this particle  polarization in, for instance, the SM process $e^+e^-\to Z H$, 
with $H$ being the Higgs boson and the electron and positron assumed unpolarized. The polarization matrix of the $Z$ boson is given by~\cite{Nakamura:2017ihk,Rao:2023jpr}
\be
\left\{ 
\begin{array}{ccc}
\rho_{_{00}} &\propto& 4\, m_{_{0}}  \dfrac{E_{Z}^{2}}{m_{Z}^{2} }(c_V^2+c_A^2) \sin^2 \theta\\
\rho_{_{++}} &\propto&  m_{_{0}}\Big[ (c_V+c_A)^2   (1-\cos \theta)^2+(c_V-c_A)^2  (1+\cos \theta)^2 \Big] \\
\rho_{_{--}}&\propto& m_{_{0}} \Big[(c_V+c_A)^2   (1+\cos \theta)^2+(c_V-c_A)^2  (1-\cos \theta)^2\Big] \\
\rho_{_{+-}} = \rho_{_{-+}} & \propto&2\, m_{_{0}} (c_V^2+c_A^2) \sin^2 \theta\\
\rho_{_{+0}}   = \rho_{_{0+}}  &\propto&  \sqrt{2}\, m_{_{0}} \dfrac{E_{Z}}{m_{Z} }\sin \theta \Big[ (c_V+c_A)^2   (1-\cos \theta)-(c_V-c_A)^2  (1+\cos \theta) \Big]\\
\rho_{_{-0}}  = \rho_{_{0-}} &\propto&  \sqrt{2}\, m_{_{0}}  \dfrac{E_{Z}}{m_{Z} } \sin \theta \Big[ (c_V+c_A)^2   (1+\cos \theta)-(c_V-c_A)^2  (1-\cos \theta) \Big]
\end{array}
\right. \label{eq:Zpol}
\ee
in which $c_V=1/2\, (-1 + 4 \sin^2 \theta_W)$, $\theta_W$ is the Weinberg angle, $c_A=-1/2$,  $m_{Z}$ and $E_{Z}=(s -m_{Z}^{2} - m_{H}^{2}/(2 \sqrt{s})$  are, respectively, the mass and energy of the  gauge boson and
\be
m_{_{0}} = \frac{g^4 m_Z^2 s}{8 \cos^4 \theta_W (s-m_Z^2)^2} \, . \label {sm}
\ee
The angle $\theta$ is the polar angle between the direction of the $Z$ boson and the momentum of the electron, which is taken to be along the $z$ axis.

The coefficient $m_{_{0}}$ simplifies once the density matrix is normalized and the only  uncertainties come from the Weinberg angle, which enters the definition of $c_V$, and the mass and energy of the $Z$-boson.

Figure~\ref{fig:context_Z} shows the values of  $\overline{\mathbb{CNTXT}}_5$ and $\overline{\mathbb{CNTXT}}_9$ as we vary the the energy $E_{Z}$ of the $Z$ boson and the angle $\theta$.

Let us take  as a benchmark the $Z$ boson at rest ($E_Z\simeq m_Z$). The polarization matrix does not depend on the angle $\theta$ (see Fig.~\ref{fig:context_Z}).   We compute the context observables in \eq{eq:max} and find: 
\be
\overline{\mathbb{CNTXT}}_5=1.8726 \pm 0.0001 \quad \text{and} \quad \overline{\mathbb{CNTXT}}_9 =3.19147\pm 0.00005 \, . \label{Zcont}
\ee
Non-contextuality is violated with a significance of better than $5\sigma$ using nine projectors, while it is not violated with the test using five. %The (theoretical) uncertainties are negligible because they only come from those of the Weinberg angle.

%%%%%%%%%%%%%%%%%%%%%%%%%%%%%%%%%%%%%%%%%%%
\subsection{$W$ and $Z$ gauge bosons in  Higgs boson decays}
%%%%%%%%%%%%%%%%%%%%%%%%%%%%%%%%%%%%%%%%%%%%%%%%%%
The polarization of the massive gauge bosons can also be studied by analyzing the decays of the Higgs boson, which provide a paradigmatic example of the interplay between bipartite entanglement, individual particles polarization and contextuality.

The polarization matrix $\rho_{VV}$ for the two vector bosons emitted in the decays of the Higgs boson $H\to ZZ^*$ and $H\to WW^*$ is given by~\cite{Fabbrichesi:2023cev}
\be
\rho_{VV} = 2 \begin{pmatrix} 
  0 & 0 & 0 & 0 & 0 & 0 & 0 & 0 & 0  \\
  0 & 0 & 0 & 0 & 0 & 0 & 0 & 0 & 0  \\
  0 & 0 &  h_{44} & 0 &  h_{16} & 0 & h_{44} & 0 & 0  \\
  0 & 0 & 0 & 0 & 0 & 0 & 0 & 0 & 0  \\
  0 & 0 &  h_{16} & 0 & 2\, h_{33} & 0 & h_{16} & 0 & 0  \\
  0 & 0 & 0 & 0 & 0 & 0 & 0 & 0 & 0  \\
  0 & 0 &  h_{44} & 0 &  h_{16} & 0 &  h_{44}& 0 & 0  \\
  0 & 0 & 0 & 0 & 0 & 0 & 0 & 0 & 0  \\
  0 & 0 & 0 & 0 & 0 & 0 & 0 & 0 & 0  \\
\end{pmatrix} \, ,
\label{rhoH}
\ee
with the condition $\Tr[\rho_{VV}]=1$ following from the relation $4(h_{33}+h_{44})=1$. The coefficients are given, at the leading order in the SM, as
\bea
&&h_{16}=\frac{f M_V^2 (-m_H^2 + (1 + f^2) M_V^2)}
    {m_H^4 - 2 (1 + f^2) m_H^2 M_V^2 + (1 + 10 f^2 + f^4) M_V^4}\, ,
    \nonumber\\
    \nonumber\\
&&h_{33}=\frac{1}{4}\, \frac{(m_H^2 - (1 + f^2) M_V^2)^2}{m_H^4 - 2 (1 + f^2) m_H^2 M_V^2 + (1 + 10 f^2 + f^4) M_V^4}\, ,
    \nonumber\\
    \nonumber\\
&&h_{44}=\frac{2 f^2 M_V^4} {m_H^4 - 2 (1 + f^2) m_H^2 M_V^2 + (1 + 10 f^2 + f^4) M_V^4}\, , \label{fghHiggs}
\eea
in which $m_H$ is the Higgs boson mass, $M_V$ that of the $V=W,\ Z$ gauge boson, while $f$ gives the fraction of the off-shell gauge boson mass.

The unpolarized squared amplitude $|\xbar{{\cal M}}_{VV}|^2$ of the process instead reads
\bea
|\xbar{{\cal M}}_{VV}|^2&=&\frac{g^2\xi_V^2}{4 f^2 M_V^2}
\Big[ m_H^4 - 2 (1 + f^2) m_H^2 M_V^2 + (1 + 10 f^2 + f^4) M_V^4\Big]\, .
\eea

The density matrix in \eq{rhoH} is idempotent
due to the  identity among the correlation coefficients 
$h_{44} = 2 \left(h_{16}^2 + 2 h_{44}^2\right)$, signaling that the final $VV^*$ state is a pure state~\cite{Aguilar-Saavedra:2022wam}. 
%%%%%%%%%%%%\
\com{The density matrix in \eq{rhoH} can then be written~\cite{Aguilar-Saavedra:2022wam}
\be
\rho_H = |\Psi_H \rangle \langle \Psi_H | \, ,
\ee
where (in the basis $|\lambda\, \lambdap\rangle = |\lambda\rangle\otimes|\lambdap\rangle$ with $\lambda,\lambdap\in\{+,0,-\}$)  
\be
|\Psi_H \rangle = \frac{1}{\sqrt{2 + \gamma^2}} \left[  |{\small +-}\rangle - \gamma \,|{\small 0\, 0}\rangle + |{\small -+}\rangle  \right]
\ee
with
\be
\gamma = 1+ \frac{m_H^2 - (1+f)^2 M_V^2}{2 f M^2_V} 
\ee
and $\gamma=1$ corresponding to the production of two gauge bosons at rest.}
%%%%%%%%%%%%%%%%%%%%%%%%

The polarization matrices for the single gauge bosons are obtained by tracing out either of the two states to obtain
\be
\rho_{W,Z} = \Tr_{_{\!W,Z}}\, \rho_{VV} =2\, \begin{pmatrix} h_{44} & 0 & 0\\ 0& 2\, h_{33}&0\\ 0&0&h_{44}
\end{pmatrix}.
\ee
The polarization matrices $\rho_{W,Z}$ depend on $f M_V$, the fraction of mass assigned to the off-shell gauge boson. As discussed in the next Section, contextuality is enhanced at smaller values of the off-shell masses. By taking the benchmark value of $f=0.1$ (which correspond to the off-shell masses $M_{W^*} \simeq 8.0$ GeV and $M_{Z^*}\simeq 9.1$ GeV, respectively) we find
\be
\overline{\mathbb{CNTXT}}_5= 2.2033 \pm 0.0002 \quad \text{and}  \quad \overline{\mathbb{CNTXT}}_9 = 3.3142\pm 0.0001  \, ,
\ee
for the $W$ boson and
\be
\overline{\mathbb{CNTXT}}_5= 2.1553 \pm 0.0006 \quad \text{and}  \quad \overline{\mathbb{CNTXT}}_9 = 3.2860 \pm 0.0003 \, , \label{eq:W1}
\ee
for the $Z$ boson. In all the cases, the significance of the non-contextuality condition is more than $5\sigma$; the related uncertainties come from those affecting the values of the Higgs and gauge boson masses. These results show the feasibility of testing non-contextuality in the decays of the Higgs boson into pairs of gauge bosons.

\subsubsection{The interplay between entanglement and contextuality in polarizations}

%%%%%%%%%%%%%
\begin{figure}[h!]
\begin{center}
\includegraphics[width=2.8in]{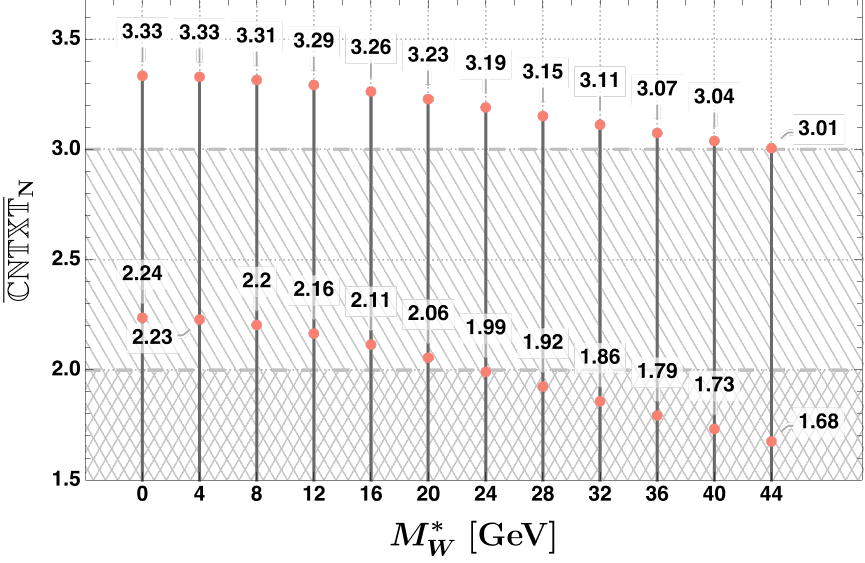}
\includegraphics[width=2.8in]{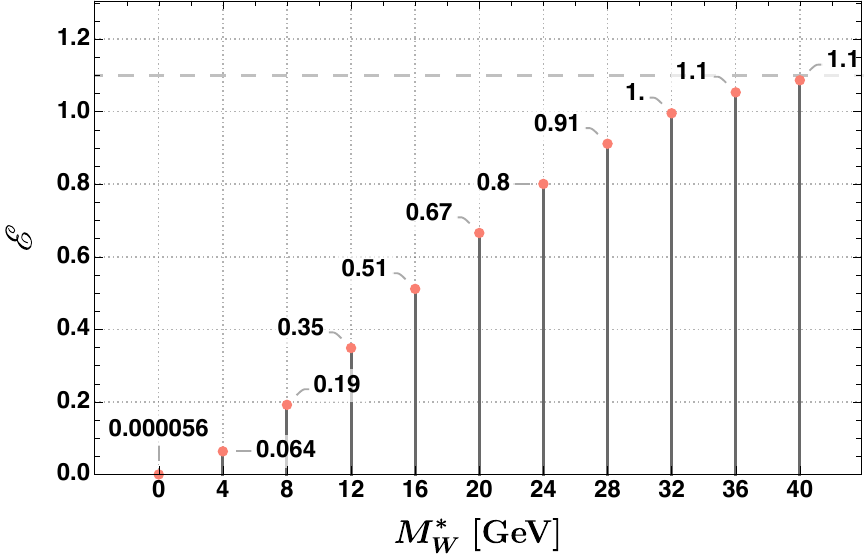}
 \caption{\footnotesize Contextuality (left-hand side) of the on-shell gauge boson $W$ as a function of $M_{W^*}$. Lower points correspond to values of $\overline{\mathbb{CNTXT}}_5$, upper points to $\overline{\mathbb{CNTXT}}_9$. Entanglement (right-hand side) of the bipartite system.
\label{fig:context_WW} 
}
\end{center}
\end{figure}
%%%%%%%%%%%% 
%%%%%%%%%%%%%
\begin{figure}[h!]
\begin{center}
\includegraphics[width=2.8in]{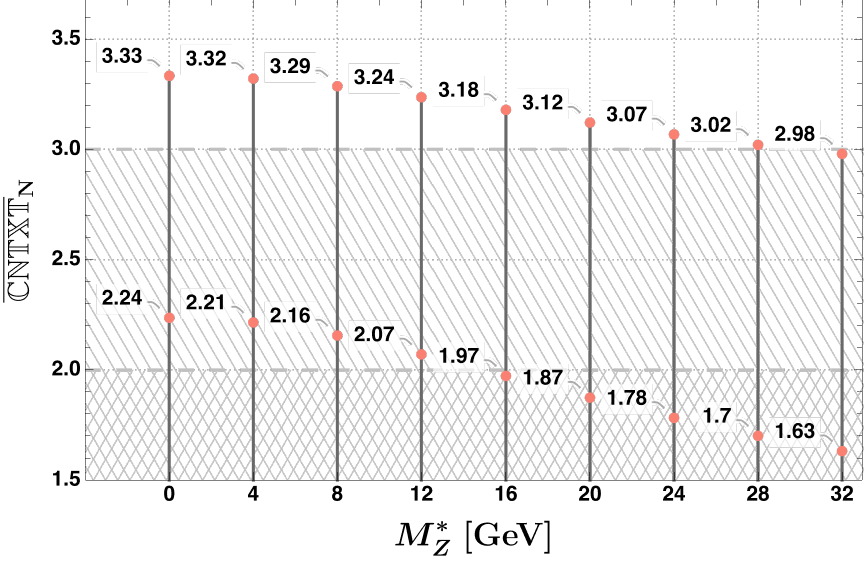}
\includegraphics[width=2.8in]{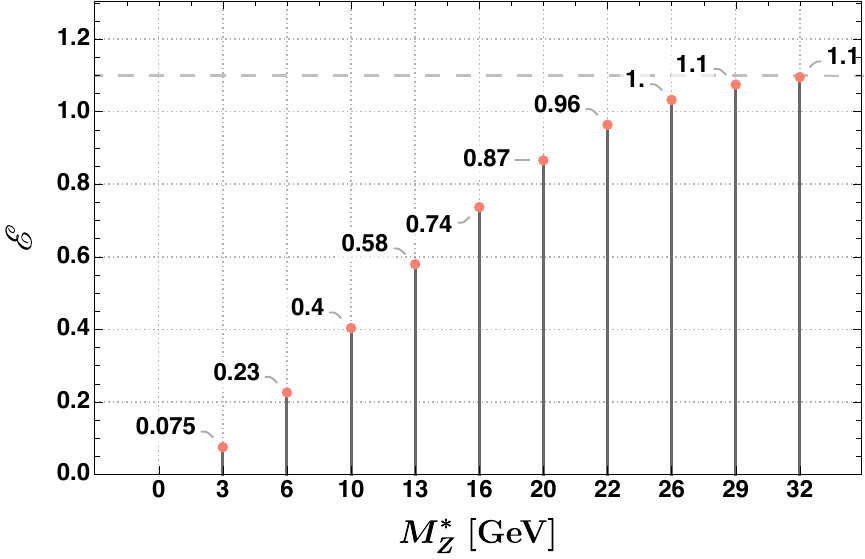}
 \caption{\footnotesize Contextuality (left-hand side) of the on-shell $Z$ gauge boson state as a function of $M_{Z^*}$. Lower points correspond to values of $\overline{\mathbb{CNTXT}}_5$, upper points to $\overline{\mathbb{CNTXT}}_9$. Entanglement (right-hand side) of the bipartite system.
\label{fig:context_ZZ} 
}
\end{center}
\end{figure}
%%%%%%%%%%%% 

If the bipartite state of interest is a pure state---like that in \eq{rhoH}---it is possible to quantify its entanglement by computing the \textbf{entropy of entanglement}: 
\be
\mathscr{E}[\rho] = - \Tr[\rho_A \log \rho_A] =  - \Tr[\rho_B \log \rho_B] \, , \label{E}
\ee
given by the von Neumann entropy~\cite{Horodecki:2009zz} of either of the two component subsystems, $A$ or $B$, described by the reduced density matrices $\rho_A={\rm Tr}_B[\rho]$ and $\rho_B={\rm Tr}_A[\rho]$.  
\com{Whereas the concurrence of a bipartite pure state is only an entanglement monotone, the von Neumann entropy is a true entanglement measure satisfying $0\leq \mathscr{E}[\rho] \leq \ln 3$, for a two-qutrit system. The first equality holds if and only if the bipartite state is separable, the second inequality saturates if the bipartite state is maximally entangled. }

It is possible to study the entanglement of the bipartite system as a function of the off-shell masses $M_{W^*}$ and $M_{Z^*}$, as shown in the right panels of Figs.~\ref{fig:context_WW} and \ref{fig:context_ZZ}. At the same time, we can plot the contextuality of the on-shell gauge bosons (left panel of the same figures).  Figs.~\ref{fig:context_WW} and \ref{fig:context_ZZ} nicely show how an increasing entanglement progressively makes the polarization state of the on-shell gauge boson more and more of an incoherent mixture and accordingly, for a certain value of the off-shell mass, the contextuality test using $\overline{\mathbb{CNTXT}}_5$ fails.

\subsection{$J/\psi$ and $\K$ mesons in $B\to J/\psi \,K^*$ decays}
%%%%%%%%%%%%%%%%%%%%%%%%%%%%%%%%%%%%%%%%%%%%%%%%%%%

Although the $J/\psi$ and $\K$ mesons can be directly produced in $pp$ collisions, their polarization is consistent with a vanishing value (for instance, see~\cite{CDF:2007msx}); larger polarizations can nevertheless be obtained when the mesons are produced through $B$-meson decays. There exists a detailed analysis of the $B^0\to J/\psi \, \K$ decay~\cite{LHCb:2013vga}, which is based on the data sample collected in $p p$ collisions at 7 TeV (part of run 1 of the LHC) with the LHCb detector for an integrated luminosity of 1 fb$^{-1}$. The branching fraction for this decay is $(1.27\pm0.05)\times 10^{-3}$~\cite{ParticleDataGroup:2022pth}.

As discussed in~\cite{LHCb:2013vga}, the selection of $B^{0}\to J/\psi \,\K$ events is based upon the analysis of the combined decays $J/\psi \to \mu^{+}\mu^{-}$ and  $\K  \to K^{+}\pi^{-}$; accordingly, their polarization state
can be reconstructed using the momenta of the final charged mesons and leptons.
The analysis in~\cite{LHCb:2013vga} gives the two complex polarization amplitudes $A_{\parallel}$ and $A_{\perp}$, as well as the non-resonant  amplitude $A_s$.
Only the former two are needed in the following, whose squared moduli and phases can be reinterpreted as follows:
\begin{align}
|A_{\parallel}|^{2} &=  0.227 \pm 0.004|_{\rm stat}\;  \pm 0.011|_{\rm sys} &
|A_{\perp}|^{2} &=  0.201 \pm 0.004|_{\rm stat}\;  \pm 0.008|_{\rm sys} \nn \\
\delta_{\parallel} \; [\text{rad}] &=  -2.94 \pm 0.02|_{\rm stat}\; \pm 0.03|_{\rm sys}  &
\delta_{\perp}  \;  [\text{rad}] &= 2.94 \pm 0.02|_{\rm stat}\; \pm 0.02|_{\rm sys}   \, ,\label{data0}
\end{align}
with $|A_{0}|^{2}+|A_{\perp}|^{2}+|A_{\parallel}|^{2}=1$, and we can take $\delta_0=0$ because there are only two independent, physical phases.
The correlations among the amplitude and phase uncertainties are also provided in~\cite{LHCb:2013vga}. 

The helicity amplitudes are mapped into the polarization amplitudes in \eq{data0} by the correspondence~\cite{Fabbrichesi:2023idl} 
\be
w_{_{00}}= A_{0} \,, \quad   w_{_{11}} = \frac{ A_{\parallel}+A_{\perp}}{\sqrt{2}}\, , \quad
 w_{_{1-1}} = \frac{ A_{\parallel} -A_{\perp}}{\sqrt{2}}\ ,
\ee
and the polarization matrix of the system $J/\psi K^*$ is then given by 
\be
\small
\rho_{J/\psi K^*} =   \begin{pmatrix} 
    w_{_{11}}  w_{_{11}}^* &0& 0 &0&  w_{_{11}}  w_{_{00}}^*& 0 & 0& 0 &  w_{_{11}} w_{_{-1-1}}^*  \\
  0 & 0 & 0 & 0 & 0 & 0 & 0 & 0 & 0  \\
 0 & 0 & 0 & 0 & 0 & 0 & 0 & 0 & 0  \\
  0 & 0 & 0 & 0 & 0 & 0 & 0 & 0 & 0  \\
   w_{_{00}} w_{_{11}}^*  & 0 &0& 0 & w_{_{00}} w_{_{00}}^*  & 0 &0& 0 &  w_{_{00}} w_{_{-1-1}}^*\\
  0 & 0 & 0 & 0 & 0 & 0 & 0 & 0 & 0  \\
  0 & 0 &  0& 0 & 0& 0 &   0& 0 & 0  \\
  0 & 0 & 0 & 0 & 0 & 0 & 0 & 0 & 0  \\
   w_{_{-1-1}} w_{_{11}}^* & 0 & 0 & 0 &   w_{_{-1-1}} w_{_{00}}^*& 0 & 0 & 0 & w_{_{-1-1}} w_{_{-1-1}}^* 
\end{pmatrix} \, .
\label{rhoBVV}
\ee
The reduced polarization matrix of the $J/\psi$ mesons is obtained from  \eq{rhoBVV} upon a partial trace over the $K^*$ meson degrees of freedom:
\be
\rho_{J/\psi}= \Tr_{\!K^*}( \rho_{J/\psi K^*} )= \begin{pmatrix} 
   |w_{_{1 1}}|^2 & 0 & 0  \\
  0 &|w_{_{0 0}}|^2  & 0   \\
  0 & 0 & &  |w_{_{-1 -1}}|^2  \\
\end{pmatrix} \, .
\label{rhoJpsi}
\ee
The contextuality of the $J/\psi$ state can now be assessed, giving:
\be
\boxed{\ \overline{\mathbb{CNTXT}}_5=1.87\pm0.01  \quad \text{and}  \quad \overline{\mathbb{CNTXT}}_9 = 3.18\pm 0.01}\ \label{eq:cntxt2}
\ee
with a violation of the non-contextuality inequality of more than  $5\sigma$ with 9 operators. No test using 5 projectors shows a violation because the original bipartite state (\ref{rhoBVV}) is highly entangled and, therefore, the reduced state (\ref{rhoJpsi}) is mostly unpolarized and less contextual---as discussed in Section 3.3.1.

Since massive spin-1  $\K$ mesons are also produced in the same decay, we can also find the contextuality for these particles by simply taking the partial trace of (\ref{rhoBVV}) over the $J/\psi$ degrees of freedom:
\be
\rho_{K^*}= \Tr_{\!J/\psi}( \rho_{J/\psi K^*} )= \begin{pmatrix} 
   |w_{_{1 1}}|^2 & 0 & 0  \\
  0 &|w_{_{0 0}}|^2  & 0   \\
  0 & 0 & &  |w_{_{-1 -1}}|^2  \\
\end{pmatrix} \, .
\label{rhochiphiphi}
\ee
Since the polarization density matrix of the $K^*$ is just that of the $J/\psi$ with two diagonal entries exchanged, we obtain the same values for the non-contextuality test as in \eq{eq:cntxt2}.

Similar results can be found by reinterpreting the data on the decays $B\to  \K \K$~\cite{LHCb:2015exo} and $B\to \phi \K$~\cite{Belle:2005lvd,LHCb:2014xzf}.

\subsection{$\phi$ mesons in $\chi^0_c \to \phi \,\phi$ and $\chi^1_c \to \phi \,\phi$ decays}
%%%%%%%%%%%%%%%%%%%%%%%%%%%%%%%%%%%%%%%%%%%%%%%%%%%

The scalar state of the charmonium, $\chi^{0}_c$, can be produced by the two-step process
\be
e^{+} e^{-} \to \psi (3686) \to \gamma \chi^{0} \, ;
\ee
it is seen to decay into a pair of $\phi$ mesons
\be
\chi^{0}_c   \to \phi + \phi \,,
\ee
with branching fraction of $(8.48\pm0.26 \pm 0.27)\times 10^{-4}$~\cite{BESIII:2023zcs}.

As the $\chi^{0}_c$ is a spin zero particle, the polarization state of the $\phi\phi$ system is a pure state, $\rho_{\phi\phi}=|\Psi\rangle\langle \Psi|$,
where, in the helicity basis, the state $|\Psi\rangle$ can be parametrized as
\be
|\Psi \rangle =w_{_{-1 -1} }\, |1,\, -1\rangle \otimes  |1,\, -1\rangle +w_{_{0 0}}\, |0\, 0 \rangle \otimes  |0\, 0 \rangle 
+  w_{_{1 1}}\, |1,\, 1\rangle \otimes  |1,\, 1\rangle  \, ,\label{pure}
\ee
with
\be
 |w_{_{-1 -1}}|^2 + |w_{_{0 0}}|^2 + |w_{_{1 1}} |^2 =1 \, ,
\ee
and $w_{_{1  1}} = -w_{_{-1 -1}}$ because of the conservation of parity. The same condition is found by considering that the final state contains indistinguishable particles. There is, therefore, only one independent amplitude and the density matrix depends on one complex number:
\be
\small
\rho_{\phi\phi} \propto  \begin{pmatrix} 
   w_{_{11}}  w_{_{11}}^* & 0 & 0 & 0 & w_{_{1 1}} w_{_{0 0}}^*& 0 & 0 & 0 &  w_{_{1 1}} w_{_{-1 -1}}^* \\
  0 & 0 & 0 & 0 & 0 & 0 & 0 & 0 & 0  \\
  0 & 0 & 0 & 0 & 0& 0 & 0& 0 & 0  \\
  0 & 0 & 0 & 0 & 0 & 0 & 0 & 0 & 0  \\
  w_{_{0 0}} w_{_{1 1}}^* & 0 & 0& 0 & w_{_{00}}  w_{_{00}}^*  & 0 & 0& 0 & w_{_{0 0}} w_{_{-1 -1}}^*  \\
  0 & 0 & 0 & 0 & 0 & 0 & 0 & 0 & 0  \\
  0 & 0 &0 & 0 & 0& 0 & 0& 0 & 0  \\
  0 & 0 & 0 & 0 & 0 & 0 & 0 & 0 & 0  \\
   w_{_{-1 -1} } w_{_{1 1}}^* & 0 & 0 & 0 &  w_{_{-1 -1}} w_{_{0 0}}^* & 0 & 0 & 0 &  w_{_{-1-1}}  w_{_{-1-1}}^*  \\
\end{pmatrix} \, ,
\label{rhochiphiphi1}
\ee
The analysis of the data in~\cite{BESIII:2023zcs}  selects $2701\pm84$ out of the $\gamma K^+K^-K^+K^-$ final states events. The maximum likelihood fit yields the absolute value of the ratio of the moduli of the helicity amplitudes:
\be
\left| \frac{w_{_{1 1}}}{w_{_{0 0}}}\right|=0.299\pm0.003|_{\rm stat} \, \pm 0.019|_{\rm syst} \, . \label{exp1}
\ee
No value for the relative phase is provided. As pointed out in~\cite{Fabbrichesi:2024rec}, this phase comes from the final state strong interactions if we assume that the form factors have no significant absorptive part. Accordingly, we can only carry out the analysis in the case of zero phase.

The polarization density matrix of either $\phi$ meson is obtained from that in Eq.~\eqref{rhochiphiphi1} upon a partial trace over the degrees of freedom of the remaining $\phi$
as
\be
\rho_\phi = \Tr_{\! \phi}( \rho_{\phi\phi}) =
  \begin{pmatrix} 
   |w_{_{1 1}}|^2 & 0 & 0  \\
  0 &|w_{_{0 0}}|^2  & 0   \\
  0 & 0 & &  |w_{_{-1 -1}}|^2  \\
\end{pmatrix} \, .
\label{rhochiphiphi}
\ee
With this result, the non-contextuality inequalities with 5 and 9 projector operators give:
\be
\boxed{\ \overline{\mathbb{CNTXT}}_5=2.11\pm0.01 \quad \text{and}  \quad \overline{\mathbb{CNTXT}}_9 = 3.26 \pm  0.01\ }
\ee
with a violation above the $5\sigma$ level in both the cases.

Data on the $\phi$ meson polarization can also be extracted from the decay of the spin-1 state $\chi^{1}_c$. Though more involved, the computation for this process follows the line of the previous one. The $\chi^{1}_c$ are produced in
\be
e^{+} e_{-} \to \psi (3686) \to \gamma \chi^{1}_c\,,
\ee
with a branching fraction of $(4.36 \pm 0.13 \pm 0.18) \times 10^{-4}$~\cite{BESIII:2023zcs}. 

There are here two independent amplitudes because the symmetry under exchange 
of identical particles in the final state gives $ w_{_{1\,1}} =  w_{_{0,0}} = 0$. The density matrix representing the polarization state of the system of two $\phi$ particles is then found to be
{\small
\be 
\rho_{\phi\phi} \propto \left(
\begin{array}{ccccccccc}
 0 & 0 & 0 & 0 & 0 & 0 & 0 & 0 & 0 \\
 0 &  |w_{1\,0}|^2(1+c^2_{\Theta})  & 0 &
    w_{1\,0} w^{\star}_{0\,1} s^2_{\Theta} & 0 &  w_{1\,0}
   w^{\star}_{0\,-1} (1+c^2_{\Theta})  & 0 &  w_{1\,0}
   w^{\star}_{-1\,0} s^2_{\Theta} & 0 \\
 0 & 0 & 0 & 0 & 0 & 0 & 0 & 0 & 0 \\
 0 &  w_{0\,1} w^{\star}_{1\,0} s^2_{\Theta} & 0 &
 |w_{0\,1}|^2(1+c^2_{\Theta}) & 0 &  w_{0\,1}
   w^{\star}_{0\,-1} s^2_{\Theta} & 0 &  w_{0\,1} w_{-1\,0}(1+c^2_{\Theta})
    & 0 \\
 0 & 0 & 0 & 0 & 0 & 0 & 0 & 0 & 0 \\
 0 &  w_{0\,-1} w^{\star}_{1\,0} (1+c^2_{\Theta}) & 0 &
 w_{0\,1} w^{\star}_{0\,-1} s^2_{\Theta} & 0 &
 |w_{0\,-1}|^2(1+c^2_{\Theta}) & 0 &  w_{0\,-1}
   w^{\star}_{-1\,0} s^2_{\Theta} & 0 \\
 0 & 0 & 0 & 0 & 0 & 0 & 0 & 0 & 0 \\
 0 &  w_{-1\,0} w^{\star}_{1\,0} s^2_{\Theta} & 0 &  w_{-1\,0}
   w^{\star}_{0\,1}  (1+c^2_{\Theta}) & 0 &  w_{-1\,0}
   w^{\star}_{0\,-1} s^2_{\Theta} & 0 &  |w_{-1\,0}|^2(1+c^2_{\Theta})
    & 0 \\
 0 & 0 & 0 & 0 & 0 & 0 & 0 & 0 & 0 \\
\end{array}
\right)\, ,\label{chi1}
\ee
}
in which $s_{\Theta}\equiv \sin \Theta$, $c_{\Theta}\equiv \cos \Theta$ and $\Theta$ is the scattering angle in the centre of mass (CM) frame.

The analysis of the data in~\cite{BESIII:2023zcs}  selects $1529\pm45$ out of the $\gamma K^+K^-K^+K^-$ final state events. The maximum likelihood fit yields the absolute value of the ratio of the moduli of the helicity amplitudes:
\be
 \left| \frac{w_{_{1\,0}}}{ w_{_{0,1}}}\right|  = 1.05\pm0.05 \quad \text{and} \quad
  \left| \frac{w_{_{1\,1}}}{ w_{_{1\,0}}} \right|= 0.07 \pm 0.04\, ,
 \ee
in which the uncertainty is only statistical. The symmetries of the decay process imply that $ w_{_{0\,1}} = - w_{_{-1,0}} $ and $w_{_{1\,0}} =  -w_{_{0\,-1}}$. The polarization matrix for a single $\phi$ meson is obtained by the reduction of the matrix in Eq.~(\ref{chi1}):
\be
\rho_\phi = \Tr_{\! \phi}( \rho_{\phi\phi}) = \frac{1}{\rho_\phi^0}
  \begin{pmatrix} 
  |w_{_{0\, -1}}|^2 (1 + c_{ \Theta}^{2})& 0 &-2\, w_{_{0\, -1}} w_{_{-1\, 0}}^{*} s_{\Theta}^{2}\\
  0 &   (|w_{_{0\, -1}}|^2 +|w_{_{-1\, 0}}|^2) (1 + c_{\Theta}^{2})& 0   \\
  -2\,  w_{_{0\, -1}} w_{_{-1\, 0}}^{*} s_{\Theta}^{2} & 0 &   |w_{_{-1\, 0}}|^2  (1 + c_{ \Theta}^{2}) \\
\end{pmatrix} \, ,
\label{rhochiphiphi}
\ee
in which $\rho_\phi^0 =2  (|w_{_{0\, -1}}|^2 +|w_{_{-1\, 0}}|^2) (1 + c_{\Theta}^{2})$.

%%%%%%%%%%%%%
\begin{figure}[h!]
\begin{center}
\includegraphics[width=2.8in]{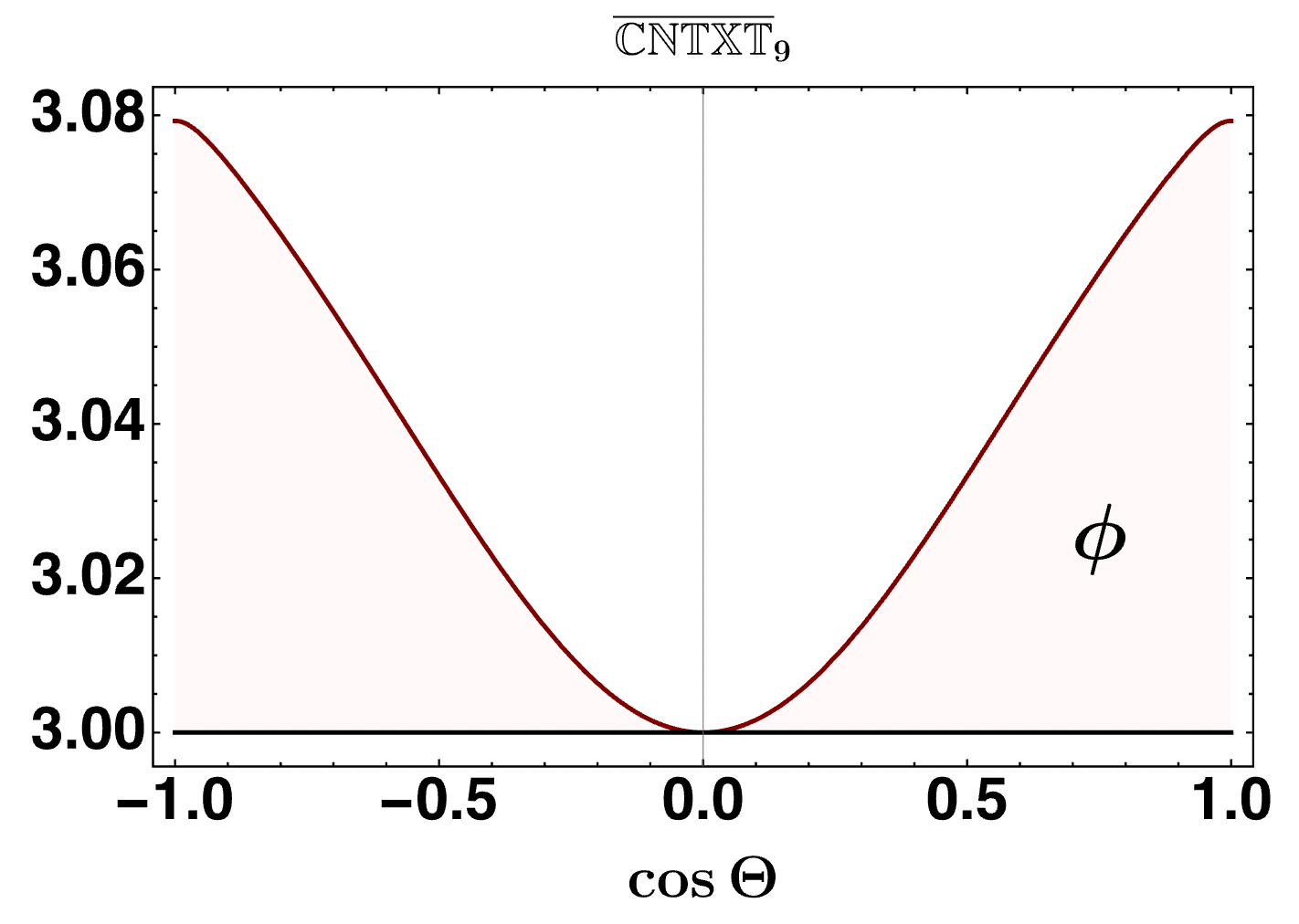}
 \caption{\footnotesize\ Contextuality  $\overline{\mathbb{CNTXT}}_9$ of the  $\phi$ spin state. $ \overline{\mathbb{CNTXT}}_5$ does not depend on the angle $\Theta$.
\label{fig:context_phi} 
}
\end{center}
\end{figure}
%%%%%%%%%%%% 
\noindent
The expectations value, after the diagonalization of the projector operators, is independent of the angle $\Theta$ for $\overline{\mathbb{CNTXT}}_5$; the angular dependence of $\overline{\mathbb{CNTXT}}_9$ is instead shown in Fig.~\ref{fig:context_phi}. For this expectation value we take as a benchmark $\Theta=\pi/4$, thus giving:
\be
\boxed{\ \overline{\mathbb{CNTXT}}_5=1.80\pm 0.01  \quad \text{and}  \quad \overline{\mathbb{CNTXT}}_9 = 3.106 \pm0.005\ }\,.
\ee
No violation is observed using 5 operators. The test using 9 operators, instead, highlights a violation of more than $5\sigma$.

\section{Testing pairs of spin-1/2 particles}
 
Bipartite systems of two spin-1/2 particles provide the next level in complexity in non-contextuality particle physics tests. We consider pairs of $\Lambda$ baryons and top quarks using the data collected at BESIII and CMS, respectively. We also consider $\tau$-lepton pairs treated with dedicated Monte Carlo simulations that reproduce the settings of SuperKEKB and of a future lepton collider working at the $Z$ peak.
 
The polarization density matrix for a system of two spin-1/2 particles can be written as
 \begin{equation}
	\label{eq:rho}
		\rho_{_{1/2\times 1/2}}= \frac{1}{4} \qty[
		\mathbb{1}_2\otimes\mathbb{1}_2 
		+ 
		\sum_i \BB_i^+ \, \qty(\sigma_i \otimes \mathbb{1}_2) 
		+ 
		\sum_j \BB_j^- \, \qty(\mathbb{1}_2 \otimes \sigma_j )
		+
		\sum_{i,j} \CC_{ij} \, \qty(\sigma_i \otimes \sigma_j)
		],
	\end{equation}
with $i,j=r, \,n, \,k$ and $\sigma_i$ being the Pauli matrices. The decomposition refers to a right-handed orthonormal basis, $\{\hn, \hr, \hk\}$ and the quantization axis for the polarization is taken along $\hk$, so that $\sigma_k\equiv\sigma_3$. In the fermion-pair center of mass frame we have
	\begin{equation}
		\hn = \frac{1}{\sin \Theta }\qty(\hp \times\hk), \quad \hr = \frac{1}{\sin \Theta }\qty(\hp-  \hk \cos \Theta)\,,
	\end{equation}
where $\hk$ is the direction of the momentum of the positively charged particle and $\Theta$ is the scattering angle. 
We take $\hp\cdot\hk = \cos\Theta$, with $\hp$ being the direction of the incoming $e^+$. 
The coefficients $\BB_i^+={\rm Tr}[\rho_{_{1/2\times 1/2}}\, (\sigma_i\otimes \mathbb{1}_2)]$ 
and $\BB_i^-={\rm Tr}[\rho_{_{1/2\times 1/2}}\, (\mathbb{1}_2\otimes\sigma_i )]$ represent single-particle polarizations, 
while the $3\times 3$ real matrix $\CC_{ij}={\rm Tr}[\rho_{_{1/2\times 1/2}}\, (\sigma_i\otimes\sigma_j)]$ gives their spin correlations. Quantum contextuality in this 4-dimensional system can be analyzed using the inequality (\ref{nambu}),
as discussed in Section 2.2.

Since we study a bipartite system, it is of interest to compare the contextuality exposed by the state (\ref{eq:rho}) with its entanglement content,
which can be quantified by means of the \textbf{concurrence}, $\mathscr{C}[\rho]$, taking values between zero (for separable, unentangled states)
and 1 (maximally entangled states). The concurrence can be analytically computed through the non-negative, auxiliary matrix
$R=\rho \,  (\sigma_y \otimes \sigma_y) \, \rho_{_{1/2\times 1/2}}^* \, (\sigma_y \otimes \sigma_y)$,
where $\rho_{_{1/2\times 1/2}}^*$ denotes the matrix with complex conjugated entries. The square roots of the eigenvalues of $R$, $r_i$, $i=1,2,3,4$, can be ordered in decreasing order and the concurrence of the state $\rho$ expressed as~\cite{Wootters:PhysRevLett.80.2245}
\begin{equation}
\mathscr{C}[\rho] = \max \big( 0, r_1-r_2-r_3-r_4 \big)\, .
\label{concurrence}
\end{equation}

 \subsection{$\Lambda\text{-}\bar \Lambda$ baryon pairs  in the decays of  the $J/\psi$}
 
{\versal We shall focus on the process}
\be
e^{+}e^{-} \to\gamma \to  c \bar c \to J/\psi \to \Lambda \bar \Lambda \, ,
\ee
where the $J/\psi$ is produced polarized, starting from unpolarized electrons and positrons. Ten billion $J/\psi$ events have been collected at the BESIII detector~\cite{BESIII:2022qax}. The decay $J/\psi \to \Lambda \bar \Lambda$ has branching fraction $(1.89 \pm 0.08) \times 10^{-3}$~\cite{ParticleDataGroup:2022pth}. The decay into $\Lambda\bar \Lambda$ pairs is reconstructed from the related dominant hadron decays: $\Lambda \to p\pi^-$ and $\bar \Lambda \to \bar p \pi^-$. 
The $J/\psi$ polarization depends on the scattering angle $\Theta$, so the correlation matrix of the two $\Lambda$ baryons depends on $\Theta$ as well.
The polarization density matrix can be expressed as~\cite{Fabbrichesi:2024rec}
\be
\small
\rho_{\Lambda\bar\Lambda} \propto 
\begin{pmatrix}  w_{_{\frac{1}{2}\, \frac{1}{2}}}  w_{_{\frac{1}{2}\, \frac{1}{2}}}^{*} s_\Theta^2 &-w_{_{\frac{1}{2}\, \frac{1}{2}}}  w_{_{\frac{1}{2}\,-\frac{1}{2}}}^{*} \frac{c_\Theta s_\Theta}{\sqrt{2}}& w_{_{\frac{1}{2}\, \frac{1}{2}}}  w_{_{-\frac{1}{2}\, \frac{1}{2}}}^{*} \frac{c_\Theta s_\Theta}{\sqrt{2}} &w_{_{\frac{1}{2}\, \frac{1}{2}}}  w_{_{-\frac{1}{2}\,-\frac{1}{2}}}^{*}  s_\Theta^2  \\
-w_{_{\frac{1}{2}\, -\frac{1}{2}}}  w_{_{\frac{1}{2}\,\frac{1}{2}}}^{*} \frac{c_\Theta s_\Theta}{\sqrt{2}}&w_{_{\frac{1}{2}\, -\frac{1}{2}}}  w_{_{\frac{1}{2}\,-\frac{1}{2}}}^{*} f_\Theta& w_{_{\frac{1}{2}\, -\frac{1}{2}}}  w_{_{-\frac{1}{2}\, \frac{1}{2}}}^{*}\frac{ s_\Theta^2}{2} &-w_{_{\frac{1}{2}\, -\frac{1}{2}}}  w_{_{-\frac{1}{2}\,-\frac{1}{2}}}^{*} \frac{c_\Theta s_\Theta}{\sqrt{2}}\\
w_{_{-\frac{1}{2}\, \frac{1}{2}}}  w_{_{\frac{1}{2}\,\frac{1}{2}}}^{*}\frac{c_\Theta s_\Theta}{\sqrt{2}}&w_{_{-\frac{1}{2}\, \frac{1}{2}}}  w_{_{\frac{1}{2}\,-\frac{1}{2}}}^{*}  \frac{ s_\Theta^2}{2} & w_{_{-\frac{1}{2}\, \frac{1}{2}}}  w_{_{-\frac{1}{2}\, \frac{1}{2}}}^{*} f_\Theta&w_{_{-\frac{1}{2}\, \frac{1}{2}}}  w_{_{-\frac{1}{2}\,-\frac{1}{2}}}^{*} \frac{c_\Theta s_\Theta}{\sqrt{2}}\\
w_{_{-\frac{1}{2}\, -\frac{1}{2}}}  w_{_{\frac{1}{2}\, \frac{1}{2}}}^{*} s_\Theta^2  &-w_{_{-\frac{1}{2}\, -\frac{1}{2}}}  w_{_{\frac{1}{2}\,-\frac{1}{2}}}^{*} \frac{c_\Theta s_\Theta}{\sqrt{2}}& w_{_{-\frac{1}{2}\, -\frac{1}{2}}}  w_{_{-\frac{1}{2}\, \frac{1}{2}}}^{*}\frac{c_\Theta s_\Theta}{\sqrt{2}}&w_{_{-\frac{1}{2}\, -\frac{1}{2}}}  w_{_{-\frac{1}{2}\,-\frac{1}{2}}}^{*} s_\Theta^2   \label{rhoLambda}
\end{pmatrix}\, ,
\ee
in which $f_\Theta \equiv (3 + \cos 2 \Theta )/4$, $s_\Theta \equiv \sin \Theta$ and $c_\Theta \equiv \cos \Theta$.

%%%%%%%%%%%%%
\begin{figure}[ht!]
\begin{center}
\includegraphics[width=2.8in]{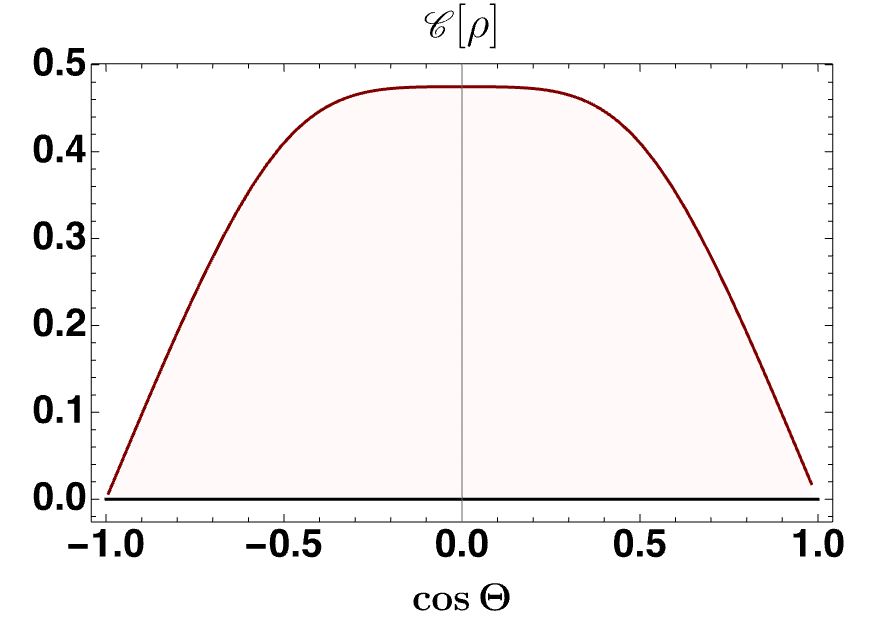}
\includegraphics[width=2.8in]{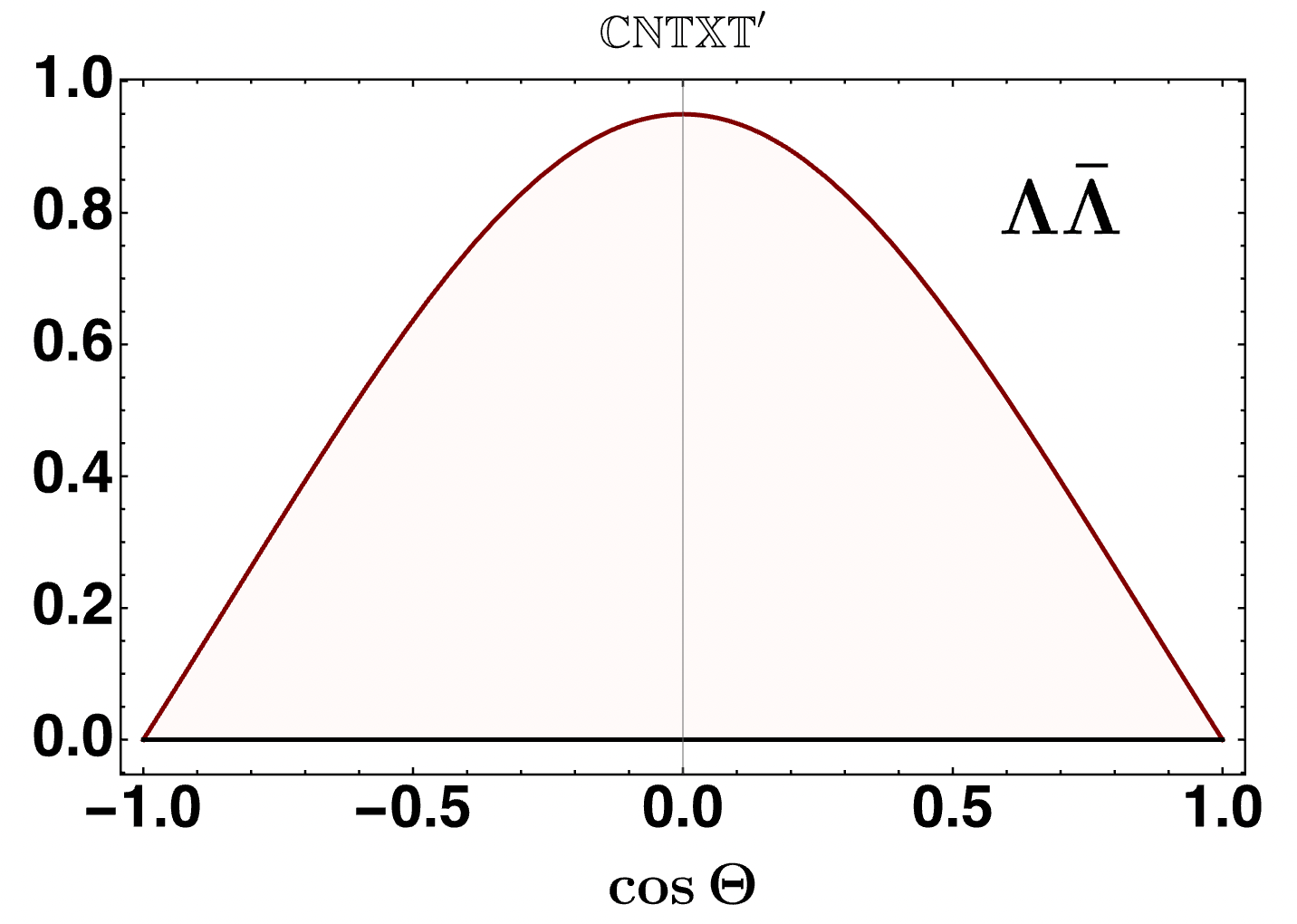}
 \caption{\footnotesize Concurrence (plot on the left-hand side) and $\mathbb{CNTXT}'$ (plot on the right-hand side) as a function of the scattering angle for the $\Lambda$-$\bar \Lambda$ pairs. The horizontal black line at 0 marks the maximal value for non-contextuality.
\label{fig:context_lambda} 
}
\end{center}
\end{figure}
%%%%%%%%%%%% 

Of the four helicity amplitudes, only two are independent; they
can be  parametrized as:
 \be
 w_{\frac{1}{2}\, \frac{1}{2}}=w_{-\frac{1}{2}\, -\frac{1}{2}}=\frac{\sqrt{1-\alpha}}{\sqrt{2}}\quad \text{ and} \quad 
  w_{\frac{1}{2}\, -\frac{1}{2}} = w_{-\frac{1}{2}\, \frac{1}{2}}=\sqrt{1+\alpha} \, \exp [-i \Delta \Phi ]\,.
  \ee
The maximum likelihood fit yields the values of the two parameters defining the helicity amplitudes~\cite{BESIII:2022qax}:
  \be
 \alpha= 0.4748 \pm 0.0022 |_{\rm stat}\pm 0.0031|_{\rm syst} \quad \text{and} \quad \Delta \Phi= 0.7521\pm0.0042|_{\rm stat} \pm 0.0066|_{\rm syst}\, .
 \ee 
No correlation in the uncertainties is provided.

We can reinterpret the analysis in terms of quantum contextuality by computing the expectation value $\mathbb{CNTXT}'$ given in Eq.~(\ref{nambu}) with the density matrix $\rho_{\Lambda\bar\Lambda}$.
The right panel of Fig.~\ref{fig:context_lambda} shows the behavior of this quantity as a function of the cosine of the scattering angle $\Theta$.
At $\Theta=\pi/2$, we have
 \be
 \boxed{\ \mathbb{CNTXT'} =0.959 \pm 0.008\ }
 \ee
and, therefore, the presence of contextuality is assessed with a significance well above $5\sigma$. For comparison, in Fig.~\ref{fig:context_lambda}, left panel, it is also reported the behavior of the concurrence, showing the strict connection between contextuality and entanglement.

A similar analysis can also be performed for the case of the decay $\psi(3686) \to \Lambda \bar \Lambda$~\cite{BESIII:2023euh},  yielding comparable results.

 \subsection{$\Sigma^-$-$ \bar \Sigma^+$ baryon pairs in the decays of  the $\psi(3686)$}

 %%%%%%%%%%%%%
\begin{figure}[ht!]
\begin{center}
\includegraphics[width=2.5in]{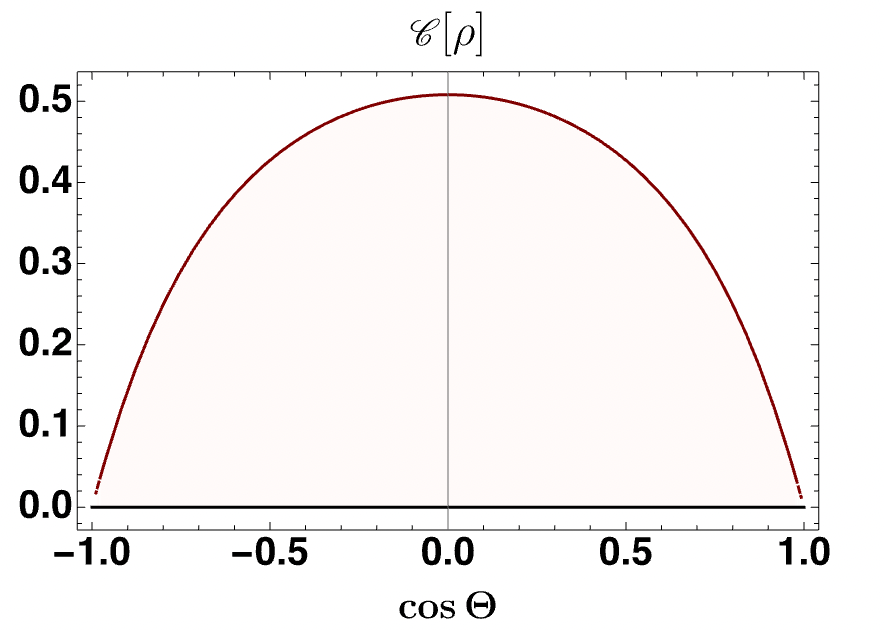}
\includegraphics[width=2.5in]{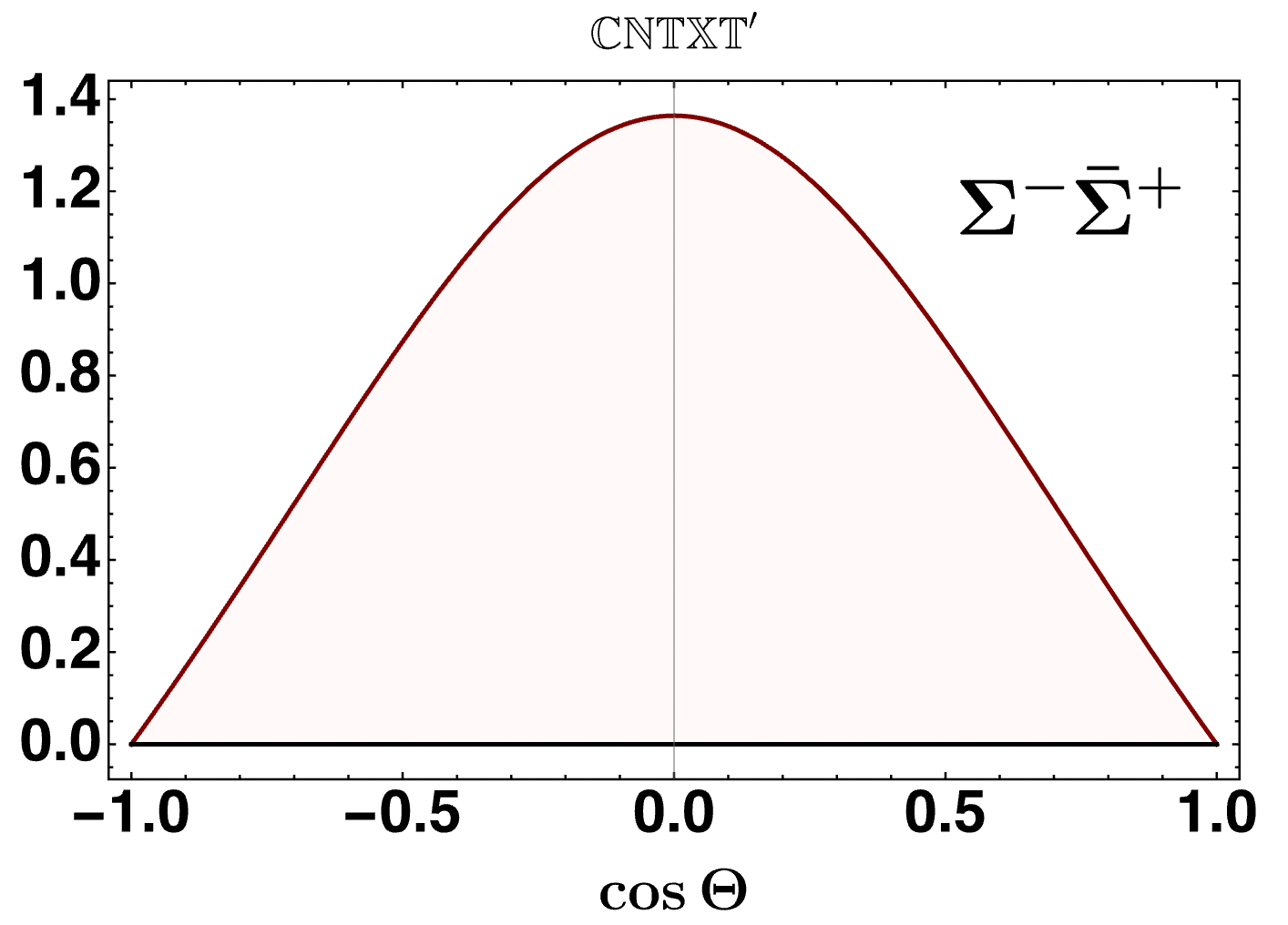}
\caption{\footnotesize Concurrence (plot on the left-hand side) and $\mathbb{CNTXT}'$ (plot on the right-hand side) as a function of the scattering angle for the $\Sigma^+$-$\bar \Sigma^-$ pairs. The horizontal black line at 0 marks the maximal value for non-contextuality.
\label{fig:sigmas} 
}
\end{center}
\end{figure}
%%%%%%%%%%%%

Similarly to the previous case, also in the process
\be
e^{+}e^{-} \to\gamma \to  c \bar c \to \psi (3686)  \to \Sigma^- + \bar \Sigma^+\, ,
\ee
the $\psi (3686)$ is produced polarized and the correlation matrix of the two baryons depends on the scattering angle $\Theta$ because the polarization of the $ \psi (3686)$ does. 

The decay  $\psi(3686) \to \Sigma^-  \bar \Sigma^+$, has branching fraction $(2.82\pm 0.09) \times 10^{-4}$~\cite{ParticleDataGroup:2022pth}. The decays $ \Sigma^- \to \bar  p\pi^0$ and  $ \bar \Sigma^+ \to  p \pi^+$ have been reconstructed out of a sample of $4.48\times 10^{8}$ $\psi(3686) $. A maximum likelihood fit over the kinematic variables yields the values of the parameters defining the helicity amplitudes~\cite{BESIII:2020fqg}:
\be
\alpha = 0.682 \pm 0.030 |_{\rm stat}\pm 0.011|_{\rm syst} \quad \text{and} \quad \Delta \Phi= 0.379\pm0.07|_{\rm stat} \pm 0.014|_{\rm syst} \, .
\ee

Again, we can reinterpret the analysis in terms of quantum contextuality by computing the expectation value  $\mathbb{CNTXT'}$ on the density matrix for the process, which structure matches that of $\rho_{\Lambda\bar\Lambda}$.
Figure~\ref{fig:sigmas} shows  the behavior  of $\mathbb{CNTXT}'$  as a function of the cosine of the scattering angle $\Theta$, together, for comparison, with that of the concurrence; here again the interplay between
quantum contextuality and entanglement is evident.
At $\Theta=\pi/2$, we have
 \be
 \boxed{\ \mathbb{CNTXT'} =1.36 \pm 0.02\ }
 \ee
and the non-contextuality inequality is violated with a significance well above $5\sigma$.

Similar results can be obtained by considering the decay $J/\psi \to \Sigma^-  \bar \Sigma^+$~\cite{BESIII:2020fqg}.

%%%%%%%%%%%%%%%%%%%%%%%%%%%%%%%%%%%%%%%%%%%
\subsection{Top quark pairs at the LHC}
%%%%%%%%%%%%%%%%%%%%%%%%%%%%%%%%%%%%%%%%%%%%%%%%%%%

The top quark is the heaviest fermion in the SM. For this reason alone a detailed study of its property is of great interest. Entanglement and Bell inequality violation have been studied for the bipartite system formed by top quark pairs~\cite{Afik:2020onf,Fabbrichesi:2021npl,Severi:2021cnj,Aguilar-Saavedra:2022uye,Dong:2023xiw,Han:2023fci}. The presence of entanglement between the top-quark pairs has also been recently confirmed by the LHC experimental collaborations~\cite{ATLAS:2023fsd,CMS:2024zkc}.

Pairs of top quarks are produced at the LHC in proton-proton collisions and their polarization can be estimated by means of the angular dependence of the momenta of suitable decay products, both for the fully leptonic and semi-leptonic decays.

The cross section for the process, summing over the top pair spins, is given by
\be
\frac{\di \sigma}{\di \Omega\, \di m_{t\bar t}} = \frac{\alpha_s^2 \beta_t}{64 \pi^2  m_{t\bar t}^2} \Big\{ L^{gg} (\tau) \, \tilde A^{gg}[m_{t\bar t},\, \Theta]+L^{qq} (\tau)\, \tilde A^{qq}[m_{t\bar t},\, \Theta]  \Big\} \label{x-sec-tt}
\ee
where $\tau = m_{t\bar t}/\sqrt{s}$, $\beta_t=\sqrt{1-4\,m_t/m_{t\bar t}}$, $\alpha_s=g^2/4 \pi$,  $m_t$ is the top mass and with $m_{t\bar t}$ being the invariant mass of the top pair. The quantities $\tilde A^{qq}$ and $\tilde A^{gg}$ are given in Ref.~\cite{Fabbrichesi:2022ovb}. The combination of the two channels  $g+g\to t +\bar t$ and $q + \bar q \to t +\bar t$ in \eq{x-sec-tt} is weighted by the respective parton luminosity functions
\be
L^{gg} (\tau)= \frac{2 \tau}{\sqrt{s}} \int_\tau^{1/\tau} \frac{\di z}{z} q_{g} (\tau z) q_{g} \left( \frac{\tau}{z}\right)\quad \text{and}\quad
L^{qq} (\tau)= \sum_{q=u,d,s}\frac{4 \tau}{\sqrt{s}} \int_\tau^{1/\tau} \frac{\di z}{z} q_{q} (\tau z) q_{\bar q} \left( \frac{\tau}{z}\right)\, ,
\ee
where we indicated with $q_j(x)$ the PDFs. For their evaluation we use the recent set ({\sc PDF4LHC21})~\cite{PDF4LHCWorkingGroup:2022cjn} setting $\sqrt{s}=13$ TeV and the factorization scale $q_0=m_{t\bar t}$ (we have used for all our results the subset 40).

The correlation coefficients entering the polarization matrix are given as
\be
\CC_{ij} [m_{t\bar t},\, \Theta]= \frac{L^{gg} (\tau)\, \tilde C_{ij}^{gg}[m_{t\bar t},\, \Theta]+L^{qq} (\tau)\, \tilde C_{ij}^{qq}[m_{t\bar t},\, \Theta]} {L^{gg}(\tau) \, \tilde A^{gg}[m_{t\bar t},\, \Theta]+L^{qq} (\tau)\, \tilde A^{qq}[m_{t\bar t},\, \Theta]} \, ,\label{pdf}
\ee
in which the values of the $\tilde C_{ij}$ coefficients were first computed in~\cite{Bernreuther:2010ny}. The explicit expressions we use can be found in \cite{Fabbrichesi:2022ovb}. Next-to-leading order (NLO) corrections where computed in \cite{Czakon:2020qbd} (NNLO QCD) and \cite{Frederix:2021zsh} (NLO EW+QCD). They mostly affect the smaller off-diagonal correlation coefficients. We neglect them in ours analytic computation, which is only of a qualitative nature.

%%%%%%%%%%%%%
\begin{figure}[ht!]
  \begin{center}
  \includegraphics[width=0.47\linewidth]{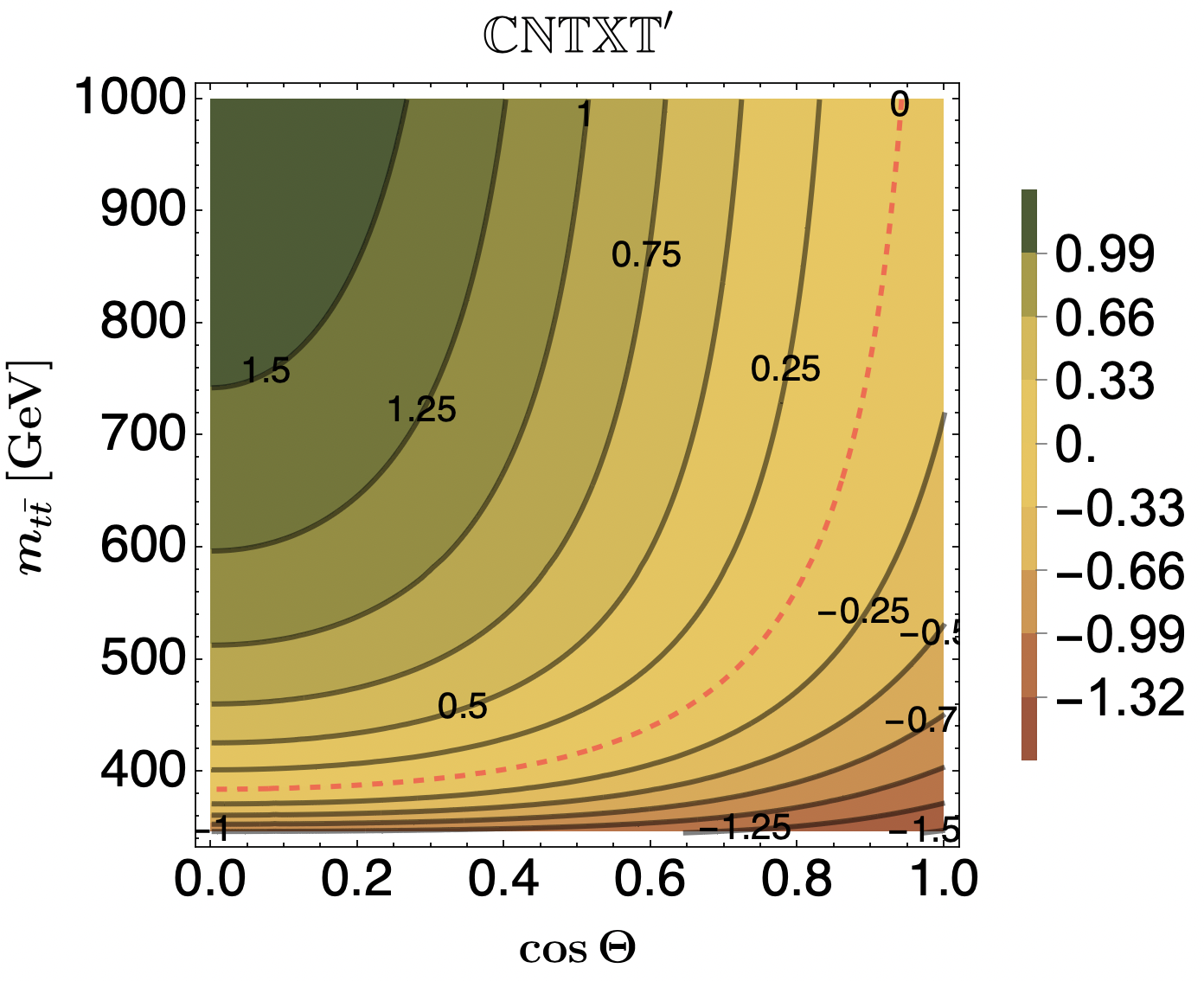} \hskip1em
  \includegraphics[width=0.46\linewidth]{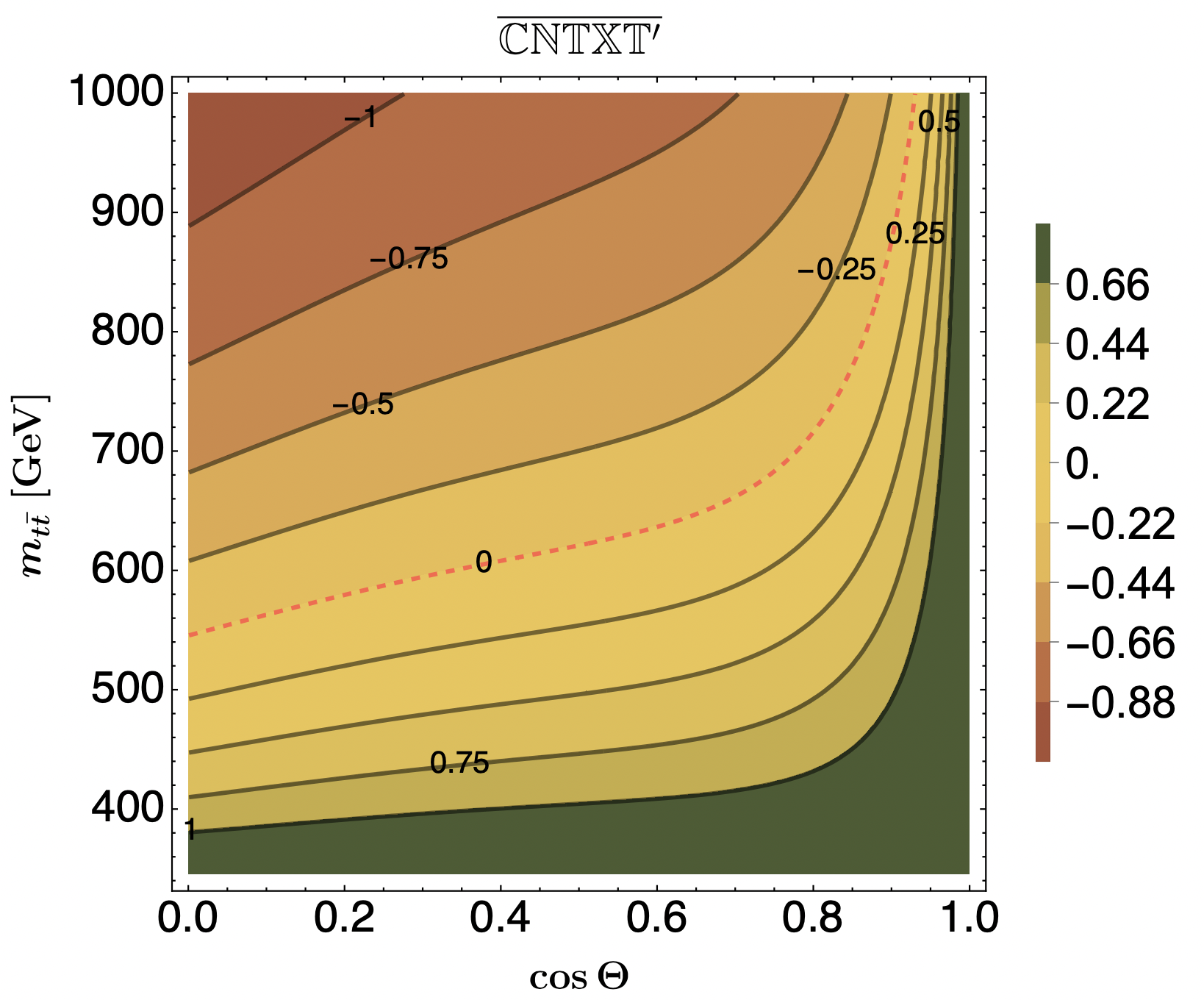}
   \caption{\footnotesize %Concurrence  $\mathscr{C}[\rho]$ (left-hand side) and  
   $\mathbb{CNTXT'}$ (left-hand side) and $\overline{\mathbb{CNTXT'}}$ (right-hand side)  as function of the invariant mass 
   $m_{t\bar t}$ and the scattering angle $\Theta$ for the process $p+ p \to t+\bar t$. The red, dashed lines mark the vanishing of the expectation values.
  \label{fig:context_tops} 
  }
  \end{center}
  \end{figure}  

Whereas top-quark pair spin states show contextuality by means of $\mathbb{CNTXT'}$ in a large section of  the accessible kinematic  space, the criterium in \eq{nambu} fails in the region below the red, dashed curve in the left panel of Fig.~\ref{fig:context_tops}. Therefore, in that region we use a more optimized operator 
\be
\overline{\mathbb{CNTXT'}}\equiv \left\langle
(U \otimes V)^{\dag}\cdot \Big[(\sigma_3\otimes \sigma_1)(\sigma_1 \otimes \sigma_3)-
(\sigma_3 \otimes \sigma_3)(\sigma_1 \otimes \sigma_1)\Big] \cdot (U \otimes V) \right\rangle \, , \label{eq:maxFF}
\ee
which gives the positive values shown in the right panel of Fig.~\ref{fig:context_tops}. The matrices selected by the optimization process are:
\be
U= \left(\begin{array}{cc}-\dfrac{2}{13}-\dfrac{48 i}{49} & \dfrac{1}{13}+\dfrac{i}{33} \\[+1em]
\dfrac{1}{19}+\dfrac{i}{16} & \dfrac{40}{41}-\dfrac{i}{7} \end{array}\right) \quad \text{and} \quad V=\left(\begin{array}{cc}\dfrac{8}{11}-\dfrac{13 i}{19} & \dfrac{1}{58}+\dfrac{i}{12} \\[+1em]
\dfrac{1}{43}-\dfrac{i}{12} & -\dfrac{2}{3}-\dfrac{8 i}{11} \end{array}\right)
\label{eqUVmax}.
\ee

\begin{figure}[h!]
  \begin{center}
  \includegraphics[width=0.5\linewidth]{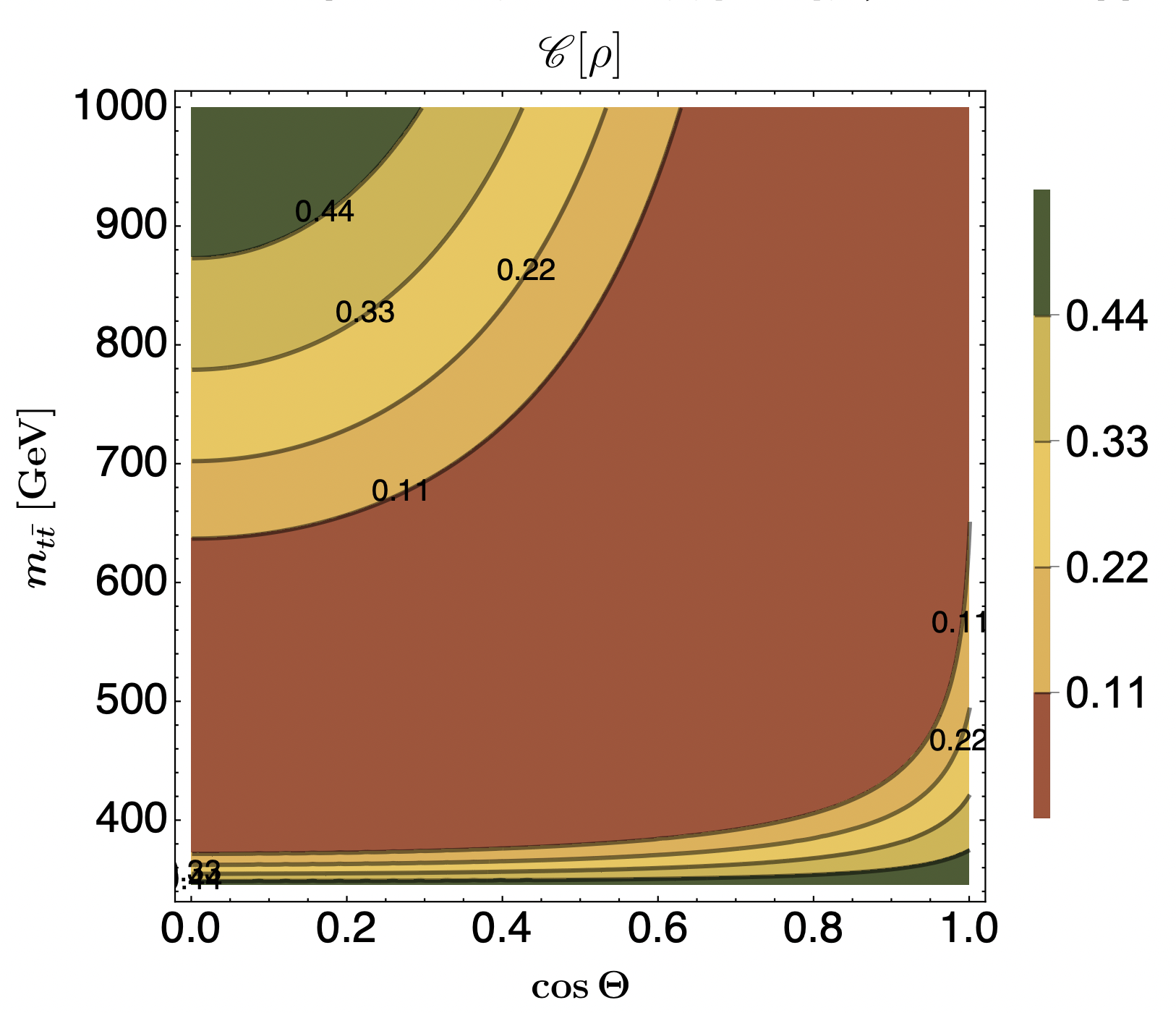}
  \includegraphics[width=0.465\linewidth]{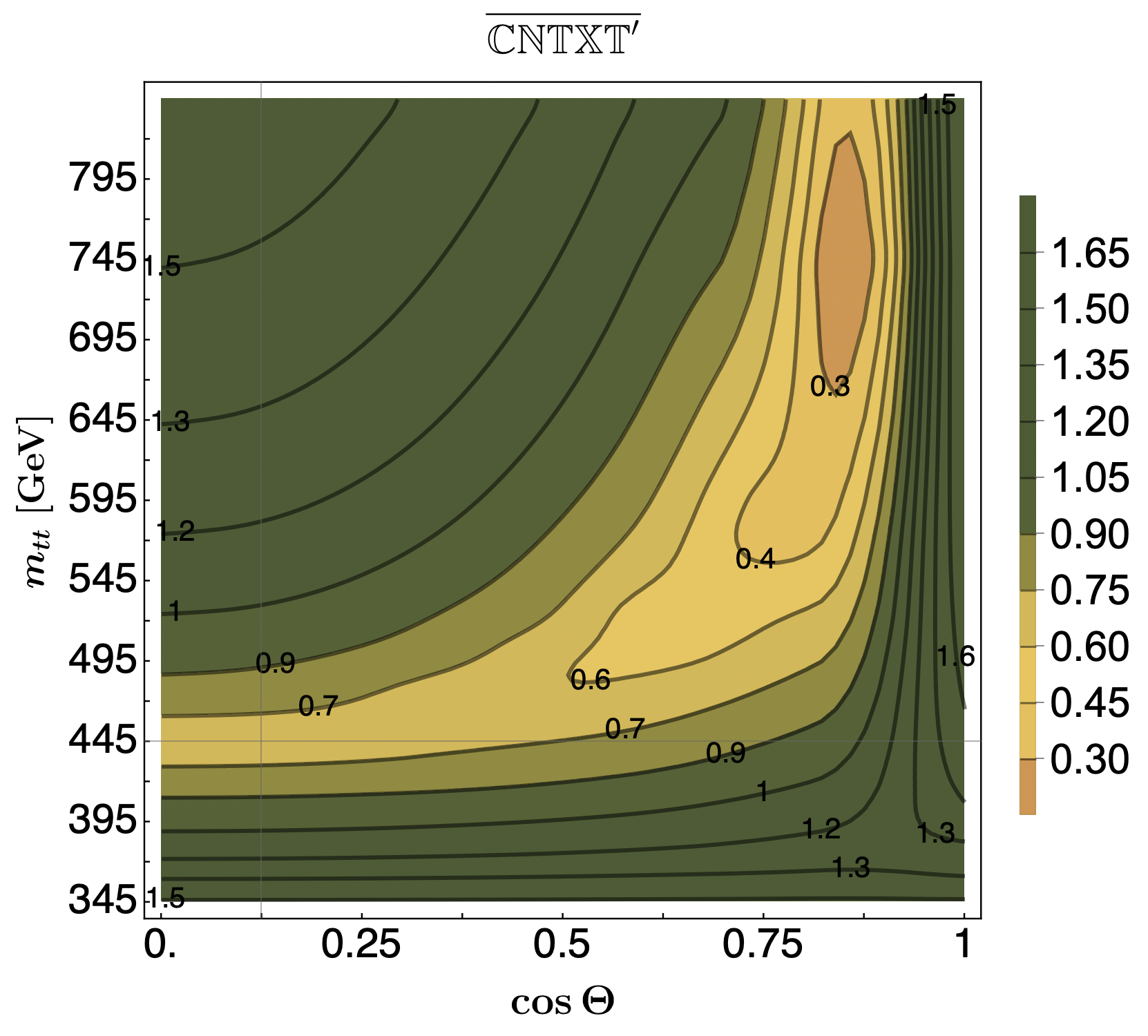}
   \caption{\footnotesize Concurrence  $\mathscr{C}[\rho]$ (left-hand side) and  
   $\overline{\mathbb{CNTXT}'}$ (left-hand side, see main text for definition)   as function of the invariant mass 
   $m_{t\bar t}$ and the scattering angle $\Theta$ for the process $p+ p \to t+\bar t$. \label{fig:c_tops} 
  }
  \end{center}
  \end{figure}

For the sake of reference, the left panel of Fig.~\ref{fig:c_tops} shows the concurrence $\mathscr{C}[\rho]$ for the top quark pairs as the invariant mass $m_{t\bar t}$ and scattering angle $\Theta$ are varied. The plot on the right-hand side shows instead the value of $\overline{\mathbb{CNTXT'}}$ obtained by optimizing the observable in each point of the considered kinematic space. Because we only need the non-contextuality  inequality to be violated even if not maximally, in our analysis we considered the matrix $U$ and $V$ which optimize the expectation value in \eq{eq:maxFF} near the top-quark pair production threshold---see \eq{eqUVmax} and the right panel of Fig.~\ref{fig:context_tops}.

%%%%%%%%%%%%%

Taking as guidance the features we have found in the analytic computations,  we want to investigate the non-contextuality inequality in \eq{nambu} and \eq{eq:maxFF} with actual LHC data---which we do in the next Section.

\subsubsection{Reinterpreting the CMS data analysis}

A recent analysis of the CMS collaboration~\cite{CMS:2024zkc} presents the measurements of the polarization 
correlations in top quark pairs using proton-proton collisions at a
CM energy of 13 TeV. The analysis is performed on data collected by the
CMS detector between 2016 and 2018, corresponding to an
integrated luminosity of 138 fb$^{-1}$.  The study is based on final states containing two $b$ jets, two jets from one $W$ boson, and an electron or muon paired together with a neutrino from the other $W$ boson.  The values of the spin correlations $\CC_{ij}$ of $t\bar t$ pairs are reconstructed for different bins in the kinematic space. Three of these are shown in Table \ref{tab:CMS}: in the first considered bin the pair of top quarks is almost at rest, while the other two correspond to a boosted region and differ only by a cut on the CM scattering angle. We use these bins to investigate quantum contextuality.

%%%%%%%%%%%%%%%%%%%%%%%%%%%%%%%%%%%
\begin{table}[h!]
\tablestyle[sansboldbw]
\begin{tabular}{*{4}{p{0.2\textwidth}}}
\theadstart
 \thead &   \thead threshold   &\thead  boosted &
    \thead  central boosted\\
\tbody
$\CC_{nn}$  & $0.540 \pm 0.042 $  & $0.175\pm0.028$ &$0.661\pm0.064$  \\
$\CC_{rr}$    &  $0.269\pm 0.070$&$-0.202\pm0.044$ & $-0.678\pm0.083$  \\
 $\CC_{kk} $    & $0.427 \pm 0.074$  &$0.040\pm0.050$  & $-0.69\pm0.12$ \\
 $\CC_{nr}$    & $0.01  \pm 0.08$  &$-0.03\pm0.05$   &$0.05\pm0.10$  \\
 $\CC_{nk}$   & $0.07 \pm 0.12$  & $-0.05\pm0.05$ & $-0.12\pm0.14$\\
 $\CC_{rn}$    & $0.07 \pm 0.08$ &$-0.04\pm0.05$  & $0.09\pm0.10$ \\
 $\CC_{rk}$     & $-0.01 \pm 0.14$ &  $-0.05\pm0.07$ & $-0.01\pm0.15$ \\
  $\CC_{kn}$   & $0.00 \pm 0.12$   & $-0.02\pm0.05$ & $-0.06\pm0.14$\\
  $\CC_{kr}$     &  $ 0.05 \pm 0.14$  &$-0.06\pm0.07$   &$0.01\pm0.15$ \\
  \hline%
  $\BB_n^+$   & $0.015\pm0.029$  & $0.007\pm0.017$ &$0.004\pm0.027$ \\
  $\BB_r^+$    & $0.004\pm0.034$  &$0.006\pm0.014$  & $-0.026\pm0.029$\\
$\BB_k^+$     & $0.001\pm0.022$ &  $0.000\pm0.011$ &$-0.015\pm0.034$\\
 $\BB_n^-$   & $-0.012\pm0.029$  & $-0.003\pm0.011$ & $-0.004\pm0.027$\\
 $\BB_r^-$    & $-0.063\pm0.033$  &$-0.013\pm0.014$ &$-0.010\pm0.029$  \\
$\BB_k^-$     & $0.003\pm0.22$ &  $0.026\pm0.017$ & $0.002\pm0.034$\\
 \hline%
  \tend
\end{tabular}
\caption{\label{tab:CMS} \footnotesize \label{tab:CijCMS} \textrm{Input coefficients for the correlations $\CC_{ij}$ and polarizations $\BB^\pm_i$ in the  three bins considered in the main text:  \underline{Threshold}: $300 < m_{tt} < 400$ GeV, \underline{Boosted}: $m_{tt} > 800$ GeV and  \underline{Central boosted}: $m_{tt} > 800$ GeV, $|\cos \Theta|<0.4$. All data from \cite{CMS:2024zkc}.}}
\end{table}
%%%%%%%%%%%%%%%%%%%%%%%%%%%%%%%%

We can first use the  values found in  the kinematic region $m_{tt}>800$ GeV and $|\cos \Theta| < 0.4$ as a benchmark to estimate the non-contextuality inequality in \eq{nambu}.
We find
\be
\boxed{\ \mathbb{CNTXT'}= 1.35 \pm 0.17 \ }
\label{cntxt-top1}
\ee
with a violation of the non-contextuality of more than $5\sigma$. This result is in agreement with what found from the analytic computation. For the bin which includes the full range of values of the scattering angle we find
\be
\boxed{\ \mathbb{CNTXT'}= 0.40 \pm 0.09\ }
\label{cntxt-top2}
\ee
in agreement, again, with the analytic computation.

The estimate of the inequality in \eq{eq:max} near the threshold region, $300 < m_{tt} < 400$ GeV, yields
 \be
\boxed{\ \overline{\mathbb{CNTXT'}}= 1.09\pm 0.09 \ }
\label{cntxt-top3}
\ee
with a violation of the non-contextuality of more than $5\sigma$.   
Errors correlations are not included in the evaluation of the uncertainties in Eqs.~(\ref{cntxt-top1})--(\ref{cntxt-top3}).

These results allow to conclude that the contextuality of top-quark pair polarization states has been established with LHC data.

\subsection{$\tau$-lepton pairs at SuperKEKB and future lepton colliders}

The $\tau$-lepton pair production at lepton colliders
\be
 \ell^{+}+ \ell^{-}  \to \tau^{+}+\tau^{-}
\ee
provides the ideal laboratory for a study of quantum mechanics at colliders, as the polarizations of the two $\tau$ leptons can be well reconstructed for several of the final states generated by the subsequent decay of these particles. Entanglement and Bell inequality violation has been extensively studied with this system~\cite{Fabbrichesi:2022ovb,Ehataht:2023zzt,Fabbrichesi:2024xtq,Fabbrichesi:2024wcd,Han:2025ewp,Fabbrichesi:2025ywl,Zhang:2025mmm}.
 
 %%%%%%%%%%%%%
\begin{figure}[h!]
\begin{center}
\includegraphics[width=0.495\linewidth]{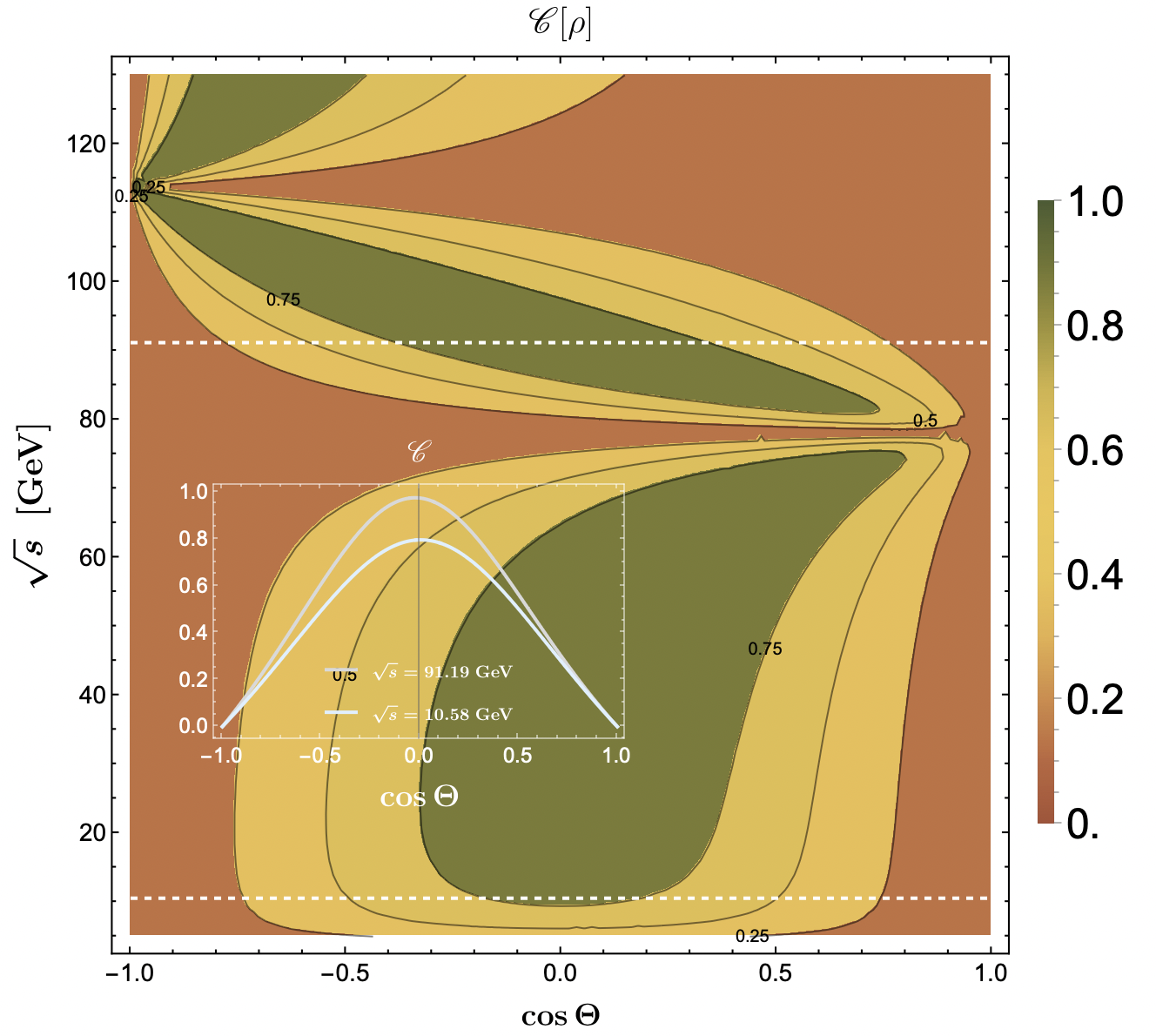}
\includegraphics[width=0.495\linewidth]{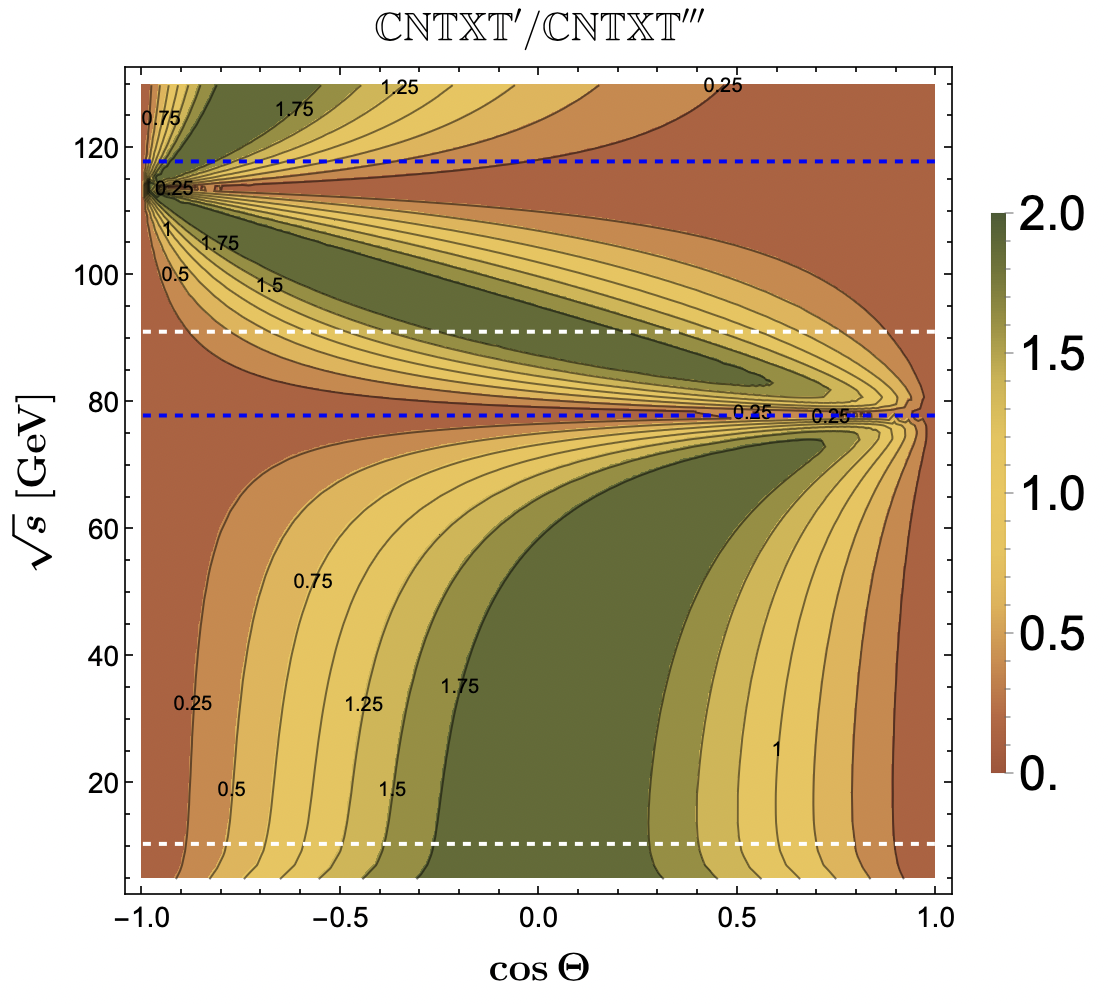}
\caption{\footnotesize Concurrence $\mathscr{C}[\rho]$ and $\mathbb{CNTXT'}/\mathbb{CNTXT'''}$ and as functions of the CM energy,  $\sqrt{s}$, and of the CM scattering angle, $\Theta$, for the process $\ell^{+} +\ell^{-}\to \tau+\bar \tau$. The energies of SuperKEK and the $Z$ boson peak are shown by the white dashed horizontal lines. The angular dependence of the concurrence at these two energies is shown by the inlet in the plot on the left-hand side.
\label{fig:plots_taus2} 
}
\end{center}
\end{figure}
%%%%%%%%%%%%%%%%%%%%%%%%%%%%%%%%%%%%%%%%%%

Let us first see what to expect by means of analytic results for the amplitudes and the correlation coefficients.  Analytic results can be obtained by means of the  expressions given in the Appendix of~\cite{Fabbrichesi:2022ovb} 
for the polarization $\BB_i^\pm$ and correlation $\CC_{ij}$ coefficients, after implementing the appropriate changes in the charge of the particles. The polarization matrix $\rho_{\tau\bar \tau}$ can thus be computed and the expectation value of the operator in \eq{nambu} readily obtained.

The right panel of Fig.~\ref{fig:plots_taus2} shows the values for $\mathbb{CNTXT'}$ and  $\mathbb{CNTXT'''}$ in the kinematic space of the $\tau$-lepton pair production.  The non-contextuality inequality with $\mathbb{CNTXT'}$ is violated in the lower and upper energy ranges. These are the two regions above and below the blue, dashed horizontal lines in the right panel of Fig.~\ref{fig:plots_taus2}. The non-contextuality inequality with $\mathbb{CNTXT'''}$  is, instead, violated in the energy region between the two blue, dashed lines. As it can be seen from the plot, the SuperKEKB CM energy is covered by the contextuality test performed by using $\mathbb{CNTXT'}$, while that around the $Z$ boson peak by $\mathbb{CNTXT'''}$. For the sake of reference, in left panel of Fig.~\ref{fig:plots_taus2} we show the concurrence values that quantify entanglement in the spin correlations of the $\tau$-lepton pairs.

The proposed theoretical reconnaissance has thus identified and estimated a number of features, which  we check next by means of the Monte Carlo simulations presented in the next Section.

\subsubsection{Monte Carlo simulations}

We want to study the $\tau$-lepton pair production at electron-positron colliders
 \be
 e^{+} + e^{-} \to  \gamma, Z \to \tau^{+} + \tau^{-} \to \pi^+\pi^- \nu_\tau \bar \nu_\tau\, ,
 \ee
in which, for simplicity, we only look at final states containing two pions and two neutrinos. The latter can be fully reconstructed via the available kinematic constraints supplemented with information on the tau decay vertices. In fact, the $\tau$ lepton lives long enough to give a decay vertex that can be distinguished from the collision point. Consequently, the vector of closest approach, identified from the continuation of the trajectories of the pions emitted in the decay, can be measured and used to resolve the two-fold degeneracy arising from the momenta reconstruction.
 
In our study we use existing Monte Carlo simulations for pairs of $\tau$ leptons generated at SuperKEKB (for the BelleII experiment with CM energy $\sqrt{s} = 10.579$ GeV)~\cite{Ehataht:2023zzt} and for a future lepton collider working at the $Z$ boson peak~\cite{Fabbrichesi:2024wcd}. 

In the simulations, the Monte Carlo truth $\tau$ lepton momenta are replaced with those obtained from the neutrino momenta reconstruction. To further mimic real data, the efficiency and uncertainty in the charged pion tracks and vertex determination in the detector are included prior to performing the neutrino momenta reconstruction and  the sample is contaminated  with events at lower $\sqrt s$ to include the effect of initial-state radiation (ISR).  A realistic determination of the experimental systematic uncertainties is not possible until the machine and the detectors are comprehensively understood. The size of the systematic errors affecting the simulation is partially taken into account by comparing  the absolute value of the differences between the central values  obtained in the Monte Carlo simulation without and with detector effects and ISR. We sum in quadrature the statistical and systematic errors.

The theoretical uncertainties are either very small (those coming from NLO contributions, or the uncertainty in the value of $m_Z$) or not present (those originating from different parton distribution functions, which are not utilized in the first place) because of the interactions allowing the process under consideration. The dominant background arises from the possible misconstruction of the $\tau$ decay channel. This is very small for the single pion channel, much smaller than for other decay channels, for instance that into two pions. A potentially large background comes from the  presence of electron and positrons in the final state, but it can be controlled by using the impact parameter. Backgrounds arising from the process $e^+e^-\to q \bar q$ and from other sources are even smaller. It is therefore not necessary to estimate their effect.

The events thus generated and reconstructed can be used to provide an estimate of the violation of non-contextuality inequalities in $\tau$-lepton pairs. The values of the polarization matrix coefficients indicated by the Monte Carlo simulations, averaged over the angular distribution of the events, are reported in Tab.~\ref{tab:taus} together with the corresponding errors.

\begin{table}[h!]
  \tablestyle[sansboldbw]
  \begin{tabular}{*{3}{p{0.25\textwidth}}}
  \theadstart
   \thead &   \thead $Z$-boson resonance   &\thead  Belle~II ($\sqrt s =10.579$ GeV)\\
  \tbody
  $\CC_{nn}$     & $0.4819 \pm 0.0079 $  &$-0.4129\pm 0.0033$ \\
   $\CC_{rr}$    & $-0.4784 \pm 0.0084$   &$0.5273\pm 0.0032$ \\
   $\CC_{kk} $   & $1.000\pm 0.0074$      &$0.8829 \pm 0.0028$ \\
   $\CC_{nr}$    & $-0.0073 \pm 0.0082$   &$-0.0014\pm 0.0040$     \\
   $\CC_{nk}$    & $-0.0016 \pm 0.0089$   &$0.0008\pm 0.0036$ \\
   $\CC_{rn}$    & $-0.0066 \pm 0.0082$   &$0.0007\pm 0.0029$   \\
   $\CC_{rk}$    & $0.0016 \pm 0.0070$    &$0.0024\pm 0.0030$  \\
   $\CC_{kn}$    & $-0.0002\pm 0.0080 $   &$0.0024 \pm 0.0031$ \\
   $\CC_{kr}$    &  $-0.0004\pm 0.0087$   &$0.0030 \pm 0.0032$   \\
    \hline%
    $\BB_n^+$   & $-0.0028 \pm 0.0042$  & -- \\
    $\BB_r^+$   & $-0.0001 \pm 0.0049$  & -- \\
  $\BB_k^+$     & $0.2198\pm 0.0044$    & -- \\
  $\BB_n^-$     & $-0.0039\pm 0.0048$   & --\\
  $\BB_r^-$     & $0.0017\pm 0.0049$    & --   \\
  $\BB_k^-$     & $0.2207\pm 0.0044$    & --  \\
   \hline%
    \tend
  \end{tabular}
  \caption{\footnotesize \label{tab:taus} \textrm{Values of the correlation coefficients $\CC_{ij}$ and polarizations $\BB^\pm_i$ obtained by the Monte Carlo simulation  performed in Refs.~\cite{Fabbrichesi:2024wcd, Ehataht:2023zzt}, averaging on the angular distribution of the generate events. The values for the $\BB_i^\pm$ are missing in the case of Belle II because analytically vanishing.}}
\end{table}
After propagating the related uncertainties, we find, for the Belle-II set up 
\begin{equation}
\boxed{\mathbb{CNTXT'} = 1.055 \pm 0.006}\,.
\end{equation}
At the $Z$-boson resonance---to be explored by a future collider---we find, instead:
\begin{equation}
\boxed{\mathbb{CNTXT'''} = 0.964 \pm 0.016}\,.
\end{equation} 
Both the values are in reasonable agreement with the analytic results and give indication and guidance for what actual experiments might find. Together, they show the feasibility of observing the non-contextuality of $\tau$-lepton pair spin states both at SuperKEK and at any future collider working at the $Z$-boson peak.

\section{Outlook}

{\versal As it is often the case} in testing quantum mechanics, one must be aware of possible loopholes potentially nullifying the results of a test.  Of the three loopholes  mentioned in the experimental tests of contextuality---lack of sharpness of the measurements, detection inefficiencies and incompatibility of the sequential measurements---only the one about detection efficiency is relevant in particle physics.  The other two loopholes are closed by the full knowledge of the state used in the test,  obtained  by means of quantum state tomography~\cite{Fabbrichesi:2025aqp}. The detection loophole is the only one possibly relevant in particle physics because the efficiency with which the events are recorded varies. The loophole is addressed by the fair sampling assumption, which is always in place in high-energy experiments. Accordingly, we think that the tests presented in the previous Sections extend to high energies the results previously obtained in dedicated low-energy experiments in a manner that is free from the most obvious loopholes. Overall, high-energy searches, barring tests relying on state-independent inequalities, seem to perform surprisingly well in testing  quantum contextuality.

The presence of contextuality for the spin states of massive spin-1  particles has been established by reinterpreting the experimental data, or providing an analytic result within the SM for several candidates: the massive gauge boson $W$ and $Z$, the mesons $J/\psi$, $\K$ and $\phi$. The contextuality of the spin state describing bipartite qubit systems was also established by analyzing pairs of spin-1/2 particles, either by reinterpreting the available data or by means of analytic results obtained within the SM for the $\tau$-leptons, top quarks and $\Lambda$ and $\Sigma^\pm$ baryons. 

These results complement those on entanglement and Bell locality for events at colliders (see, for instance, the review~\cite{Barr:2024djo} and the works therein cited) in the study of quantum mechanics at high energies and in the presence of strong and electroweak interactions. Quantum mechanics is confirmed and non-contextual hidden variable models ruled out with a significance of more than $5\sigma$.

%%%%%%%%%%%%%%%%%%%%%%%%%%%%%%%%%%%%%%%%%%%%%%%%%%%
%%%%%%%%%%%%%%%%%%%%%%%%%%%%%%%%%%%%%%%%%%%%%%%%%%%
%\newpage
\section*{Acknowledgements}
{\small
LM is supported by the Estonian Research Council under the RVTT3, TK202 and PRG1884 grants.}
\newpage
\begin{multicols}{2}
%\twocolumn  
\small
\bibliographystyle{JHEP}   
\bibliography{context.bib} 

\end{multicols}
\end{document}